\documentclass[a4paper, amsfonts, amssymb, amsmath, reprint, showkeys, nofootinbib, twoside, onecolumn, notitlepage]{revtex4-1}

\usepackage[centering,hmargin=2cm,vmargin=2cm,lmargin=1.7cm,rmargin=1.7cm]{geometry}

\def\mnras{{MNRAS}}             
\def\prd{{Phys.~Rev.~D}}        
\def\prl{{Phys.~Rev.~Lett.}}    
\usepackage{amsmath,amstext}
\usepackage[T1]{fontenc}
\usepackage{amssymb}
\usepackage{graphicx}
\usepackage{ae,aecompl}

\DeclareFontFamily{OT1}{pzc}{}
\DeclareFontShape{OT1}{pzc}{m}{it}{<-> s * [1.10] pzcmi7t}{}
\DeclareMathAlphabet{\mathpzc}{OT1}{pzc}{m}{it}

\usepackage{hyperref}
\usepackage{amsmath}
\usepackage{amssymb}
\usepackage{mathtools}
\usepackage{bm}
\usepackage{cleveref}
\usepackage{tensor}
\usepackage{braket}
\usepackage{enumitem}
\usepackage{mhchem}
\usepackage{amsthm}
\usepackage{nccmath}
\usepackage{color}

\newcommand{\spc}{\quad \quad \quad}

\def\be{\begin{equation}}
\def\ee{\end{equation}}
\def\beq{\begin{eqnarray}}
\def\eeq{\end{eqnarray}}
\theoremstyle{definition}

\theoremstyle{theorem}

\begin{document}
\title{Relativistic Liquids: GENERIC or EIT?}
\author{L.~Gavassino$^1$, M.~Antonelli$^{2}$}

\affiliation{
$^1$Department of Mathematics, Vanderbilt University, Nashville, TN, USA
\\
$^2$CNRS/IN2P3, Laboratoire de Physique Corpusculaire de Caen, 14050 Caen, France
}

\begin{abstract}
We study the GENERIC hydrodynamic theory for relativistic liquids formulated by \"{O}ttinger and collaborators. We use the maximum entropy principle to derive its conditions for linear stability (in an arbitrary reference frame) and for relativistic causality. In addition, we show that, in the linear regime, its field equations can be recast into a symmetric-hyperbolic form. Once rewritten in this way, the linearised field equations turn out to be a particular realization of the Israel-Stewart theory, where some of the Israel-Stewart free parameters are constrained. This also allows us to reinterpret the GENERIC framework in view of the principles of Extended Irreversible Thermodynamics (EIT) and to discuss its physical relevance to model (possibly viscoelastic) fluids.
\end{abstract} 

\maketitle

\section{Introduction}
\label{intro}

It is often implied that, among all kinds of substances, Israel-Stewart hydrodynamics is best suited for modeling dilute relativistic gases \cite{Liu1986,Salmonson1991,rezzolla_book,Stricker2019,Denicol2012Boltzmann}. 
 Indeed, Israel and Stewart themselves supported their theory appealing to relativistic kinetic theory  \cite{Stewart_1977,Israel_Stewart_1979}. Only recently\footnote{
    It is sometimes said \cite{Baier2008} that the Israel-Stewart model emerges from a second-order expansion in the gradients. However, we do not follow this approach here, because, strictly speaking, gradient-expansion hydrodynamics does not exist beyond first order \cite{KovtunStickiness2011}. The interpretation that we adopt here is that Israel-Stewart should be regarded as a quasihydrodynamic theory \cite{Grozdanov2019,GavassinoQuasi2022}.}, 
a fact that was already well-known to the community working within the framework of Extended Irreversible Thermodynamics (EIT) has been rediscovered \cite{Jou_Extended,Muller_book,Salazar2020,GavassinoFronntiers2021}:
the Israel-Stewart field equations constitute a very common (but not universal \cite{Heller2014,GavassinoQuasi2022}) kind of quasi-equilibrium evolution\footnote{
    Regarding the precise meaning of quasi-equilibrium, see the discussion in Sec. II-A of \citep{BulkGavassino}.}, 
which is observed in many different substances \cite{Grozdanov2019,GavassinoFronntiers2021}, including superfluids \citep{GavassinoKhalatnikov2022}, fluid-radiation systems \citep{GavassinoRadiazione} and neutron star matter undergoing nuclear reactions \citep{Camelio2022,CamelioSecondo2022}. 
Therefore, we may regard the Israel-Stewart theory as a ``universality class'', which encompasses materials having very different microscopic properties, but the same non-equilibrium behaviour at macroscopic scales \cite{Denicol_Relaxation_2011}. 
For this reason, it is not so surprising that, in non-relativistic rheology, the Israel-Stewart theory goes by a different name (``Maxwell model'' \cite{maxwell_1867,Nettleton1959,Roylance2001}), and it is not used to describe gases, but another type of substance \cite{Frenkel_book,BaggioliHolography2019,BAGGIOLI20201,landau7}: liquids. 
Interestingly, \citet{maxwell_1867} also had developed his model with ideal gases in mind (just like \citet{Israel_Stewart_1979}), and his formalism began to be systematically applied to liquids only later \cite{Frenkel_book,Nettleton1959,Truesdell1965}.

To understand why the Israel-Stewart theory (or the Maxwell model) works well also for liquids, let us take a closer look at the field equation for the shear stress component $\Pi_{xy}$ (we work  in the fluid's rest frame):
\begin{equation}\label{Israel1}
\tau_\eta \partial_t \Pi_{xy}+\Pi_{xy} = -2\eta \, \partial_{(x} u_{y)} \, ,
\end{equation}
where $u_j$ is the flow velocity, $\eta$ the shear viscosity, and $\tau_\eta$ the shear relaxation time. For slow periodic perturbations of frequency $\omega$, the relaxation term $\tau_\eta \partial_t \Pi_{xy}$ is negligible, and we recover the universal Navier-Stokes behaviour \cite{LindblomRelaxation1996}:
\begin{equation}\label{pixyvisc}
\Pi_{xy} \approx -2\eta \, \partial_{(x} u_{y)} \spc (\text{if }\omega\tau_\eta \ll 1) \, .
\end{equation}
On the other hand, for fast periodic perturbations, the relaxation term dominates over $\Pi_{xy}$, and equation \eqref{Israel1} simplifies to $\tau_\eta \partial_t\Pi_{xy} \approx -2\eta \, \partial_{(x} u_{y)}$. 
At this point, since the strain rate tensor $\partial_{(j} u_{k)}$ is the time-derivative of the deformation tensor $\varepsilon_{jk}$ \cite{landau7,CarterQuintana1972}, we can recover Hooke's law by integrating in time:
\begin{equation}\label{pixyelast}
\Pi_{xy} \approx -2 \dfrac{\eta}{\tau_\eta }\varepsilon_{xy} \spc (\text{if }\omega\tau_\eta \gg 1) \, .
\end{equation} 
The equation above describes an elastic medium, with shear modulus $\eta/\tau_\eta$. 
Summing up: if the Israel-Stewart theory is valid in both the slow and fast regimes ($\omega\tau_\eta \ll 1$ and $\omega\tau_\eta \gg 1$), then it should be a suitable model to describe substances which respond like fluids to slow perturbations, and like solids to fast perturbations  (a more detailed proof is provided in Appendix \ref{kjhA}). Following the pioneering work of \citet{Frenkel_book}, this is precisely how a liquid behaves. 

The evolution of a liquid element can be visualised as a sequence of (amorphous) solid-like configurations \cite{Nettleton1959,Trachenko2016}. On short timescales, a liquid element responds to an imposed deformation trying to relax back to its initial configuration (the one it had before the deformation), thus exhibiting an elastic response. Hence, provided the frequency of an imparted mechanical oscillation is sufficiently high, the mechanical response of a liquid is indistinguishable from that of an amorphous solid, and the diffusive component of the liquid motion is absent. 
On longer timescales, the molecules can make diffusive jumps which allow the liquid element to reach a new equilibrium configuration (the one after the deformation). This originates dissipation and, therefore, viscosity. 
Indeed, the use of the Maxwell model (i.e. the non-relativistic Israel-Stewart theory) for describing liquids has produced many interesting predictions \citep{Nettleton1959,Truesdell1965}, such as the existence of gapped momentum states \cite{BAGGIOLI20201}, which have been experimentally confirmed \cite{Grimsditch1989,Scarponi2004,Hosokawa_2013,Hosokawa_2015}.

Recently, \citet{Stricker2019} have suggested that, when it comes to modeling relativistic \textit{liquids}, another theory may be even more reliable than the Israel-Stewart paradigm. Such alternative theory has been developed by \"{O}ttinger and collaborators \cite{Otting0_1998,Otting1_1998,Otting2_1999,Otting3_1999} as an application of the GENERIC formalism \cite{Grmela1997}, a modern approach to non-equilibrium thermodynamics. 
Given the many successes of the GENERIC framework in modeling complex fluids (see \cite{OttingerReview2018} for a review), it is worth studying this relativistic hydrodynamic theory in detail, and see how  it compares with respect to Israel-Stewart. 

In this paper, we aim to answer rigorously three crucial questions concerning the GENERIC relativistic hydrodynamic theory for liquids (in the precise formulation given in \citep{Stricker2019}, which we will simply call ``GENERIC theory''):
\begin{itemize}
\item[(i)] Is it thermodynamically consistent, stable, and causal? This will be addressed in Sec. \ref{sec:thermo}.
\item[(ii)] Are the experimentally-confirmed limits \eqref{pixyvisc} and \eqref{pixyelast} recovered in the appropriate regimes? We will refer to this behavior to as ``viscoelasticity'', see \citep{Nettleton1959,Truesdell1965} or App. \ref{kjhA}, and we will investigate it in Sec. \ref{visco?}.    
\item[(iii)] Does it make predictions about liquid behavior which are not attainable with  an Israel-Stewart description? This will be discussed in Sec. \ref{sec:new}.
\end{itemize} 
The three above questions constitute a hierarchy of ``reliability tests'' that the GENERIC theory should pass, if we want it to be useful. If it passes test (i), then at least it is a meaningful hydrodynamic theory. If it passes also test (ii), then we can think of applying it to liquids. If it passes also test (iii), then it is superior to Israel-Stewart, and we should definitely think of implementing it in simulations.

Throughout the article, we adopt the metric signature $(-,+,+,+)$, and work in geometrical units: $c=k_B=1$. The spacetime may be curved, but it is treated as a fixed background. The labels $a,b,c,d$ are spacetime indices (they run from 0 to 3), while the labels $j,k$ are purely spacelike indices (they run from 1 to 3). The labels $A,B,C,D$ are field multi-indices, while the labels $I,J$ are charge indices. Einstein's summation convention applies to all kinds of indices. Symmetrization, $T_{(ab)}$, and anti-symmetrization, $T_{[ab]}$, come with the pre-factor $1/2$.

\section{GENERIC theory using EIT language}

As a starting point, it is instructive to express the GENERIC theory \cite{Stricker2019} by using a notation that is as close as possible to that of Extended Irreversible Thermodynamics \cite{Hishcock1983}. This will ease the comparison with the Israel-Stewart theory, and it will make the physical content of the GENERIC theory even more transparent.

\subsection{Fields of the theory}

Every hydrodynamic theory is a classical field theory. This just means that its mathematical degrees of freedom are a collection of classical fields $\varphi^A$ \citep{GavassinoFronntiers2021}. For the GENERIC theory, such fields can be taken to be the usual ``Israel-Stewart-like'' non-equilibrium degrees of freedom  in the Eckart frame \cite{Hishcock1983}:
\begin{equation}\label{choiceofFields}
\varphi^A = \{ \rho,n,u^a,q^a,\pi^{ab}  \} \, ,
\end{equation} 
where $\rho$ is the rest-frame energy density (which also includes the rest-mass part), $n$ is the rest-frame particle density, $u^a$ is the flow velocity, $q^a$ is the heat flux, and $\pi^{ab}$ is the viscous stress tensor. The components of $u^a$, $q^a$, and $\pi^{ab}$ are not all independent, but are subject to some algebraic constraints:
\begin{equation}
\label{algconstraints}
u_a u^a+1=u_a q^a= u_a \pi^{ab}=\pi^{[ab]}=0 \, .
\end{equation}
Thus, the total number of algebraic degrees of freedom is 14, as in the Israel-Stewart theory. This just means that also the GENERIC theory treats the dissipative currents as independent dynamical variables. The viscous stress tensor $\pi^{ab}$ includes both bulk and shear terms. In the Israel-Stewart theory, one usually decomposes it into irreducible pieces:
\begin{equation}\label{shearbulk}
\pi^{ab}=\Pi^{ab}+\Pi \, h^{ab}\, , \spc \text{with }\, \Pi\indices{^a _a}=0 \, ,
\end{equation}
where $\Pi^{ab}$ ($=\Pi^{ba}$) is the shear-stress tensor, while
\begin{equation}\label{proj}
 h^{ab}= g^{ab}+u^a u^b \, , \spc 
\Pi=  \dfrac{1}{3}\pi\indices{^a _a} 
\end{equation}
are respectively the rest-frame projection tensor ($h^{ab}u_b=0$, $h\indices{^a _c}h\indices{^c _b}=h\indices{^a _b}$) and the bulk-viscous stress. However, in the GENERIC theory, it is often convenient to combine bulk and shear, and to work with the total viscous stress $\pi^{ab}$.

\subsection{Local thermodynamics of the fluid}  

Following the approach of \citet{Israel_Stewart_1979}, we can use the fields $\rho$ and $n$ to ascribe to each fluid element a notion of temperature, $T(\rho,n)$, chemical potential $\mu(\rho,n)$, and pressure, $P(\rho,n)$, by using the three equations below:
\begin{equation}\label{thermo}
\begin{split}
& s=s(\rho,n) \, , \\
& ds= \dfrac{1}{T} d\rho -\dfrac{\mu}{T} dn \, , \\
& \rho+P=Ts+\mu n \, . \\
\end{split}
\end{equation}
The first equation expresses the rest-frame entropy density (for a given $\rho$ and $n$) that a fluid element \textit{would have} if it were in local thermodynamic equilibrium, i.e. when $q^a=\pi^{ab}=0$. It must be provided by microphysics and it represents a complete description of the thermodynamics of a homogeneous system made of the specific fluid under consideration \citep{Callen_book,peliti_book}. The second equation in \eqref{thermo} is the first law of thermodynamics, while the third equation is the usual Euler relation. We are using as ``primary variables'' the quantities $\rho$ and $n$ because, as we shall see, the GENERIC theory adopts the so-called \emph{Eckart} matching conditions \cite{Bemfica2019_conformal1}, meaning that $\rho$ and $n$ coincide with the physical energy and particle density, as measured in the reference frame defined by $u^a$ \cite{Israel_Stewart_1979}. 

When dealing with $T$, $\mu$ and $P$, there is an important subtlety that one should always keep in mind: out of thermodynamic equilibrium, ``temperature'', ``chemical potential'', and ``pressure'' are not uniquely defined \cite{BulkGavassino}. Hence, the quantities $T$, $\mu$, and $P$ are just some functions of $\rho$ and $n$ that we \textit{define} to be the temperature, chemical potential and pressure of the fluid element. 
Strictly speaking, one can even abandon the thermodynamic interpretation altogether, and think of them just as effective fields \cite{Kovtun2019,BemficaDNDefinitivo2020,
DoreTorrieri2022}, which reduce to the usual temperature, chemical potential and pressure only at equilibrium. Indeed, as we shall see, the entropy density $s(\rho,n)$ is not the physical entropy density of the fluid element, but only its ``equilibrium part''.

If we know the value of two thermodynamic variables, we can go back to $\rho$ and $n$ by inverting the functional dependence. This allows one to choose the independent thermodynamic variables that are most convenient for a given problem. For example, to simplify the bridge with statistical mechanics, one is free to use $T$ and $\mu$ as degrees of freedom \cite{KovtunLecture2012,BecattiniLecture2019}, and compute the other thermodynamic variables with the aid of the differential
\begin{equation}\label{dPippuz}
dP \, = \, s \, dT+n \, d\mu \, ,
\end{equation}
which follows from \eqref{thermo}. Later, it will be useful for our purposes to treat  $\{ \mathfrak{s} ,P\}$ as our independent variables ($\mathfrak{s}=s/n$ is the entropy per particle). Most of the thermodynamic identities that we will need are summarised below:
\begin{equation}\label{dhuzz}
\begin{split}
& \dfrac{\rho +P}{n}  =: \mathfrak{h}(\mathfrak{s},P) \, , \\ 
& d\mathfrak{h} = T d \mathfrak{s} + \dfrac{dP}{n} \, , \\
& H_{\mathfrak{h}} =
\begin{bmatrix}
   \dfrac{T}{c_p} & & \dfrac{T \kappa_p}{n c_p}  \\
   \\
   \dfrac{T \kappa_p}{n c_p} & & -\dfrac{1}{n^2 \mathfrak{h} c_s^2}  \\
\end{bmatrix} \, .
 \\
\end{split}
\end{equation}
Here, $\mathfrak{h}$ is the enthalpy per particle, $c_p$ is the specific heat at constant pressure, $ \kappa_p$ is the isobaric thermal expansivity, and $c_s$ is the adiabatic speed of sound. The symbol $H_{\mathfrak{h}}$ denotes the Hessian of $\mathfrak{h}$, in the variables $\{ \mathfrak{s},P \}$.

\subsection{Constitutive relations}

The effective fields $\varphi^A$ are just a mathematical tool that can be used to characterise the state of the system. However, what physicists really want to determine are the three \textit{physical} tensors of the fluid: the particle four-current $n^a$, the stress-energy tensor $T^{ab}$ and the entropy four-current $s^a$. Thus, the most important equations of a hydrodynamic theory are the formulas that allow us to express these three physical tensors in terms of the $\varphi^A$. Such formulas are called ``constitutive relations''\footnote{ More in general, a ``constitutive relation'' in effective field theory is any expression $F=F[\varphi^A]$, which allows one to determine the value of a physical observable $F$ only in terms of the chosen degrees of freedom $\varphi^A$ \cite{GavassinoFronntiers2021,Kovtun2019,Romatschke2017}. For example, if the only degree of freedom is the temperature $T$, the Fourier law $q_j=-\kappa(T)\partial_j T$ is a constitutive relation \cite{peliti_book}. On the other hand, the Cattaneo equation $\tau(T)\partial_t q_j +q_j=-\kappa(T)\partial_j T$ \cite{cattaneo1958} is not a constitutive relation, because it does not determine $q_j$ in terms of $T$ uniquely. Hence, $q_j$ becomes an additional degree of freedom, besides $T$.}. Only when the constitutive relations are assigned, the fields $\varphi^A$ acquire an unambiguous physical meaning. 

For the choice of fields \eqref{choiceofFields}, the constitutive relations of the GENERIC theory are (see Appendix \ref{AAA} for the proof)
\begin{equation}\label{ConstitutiveGeneric}
\begin{split}
n^a ={}& n u^a \, , \\
 T^{ab} ={}& \rho u^a u^b +P h^{ab}+q^a u^b+ u^a q^b+\pi^{ab}+ b_1 \, q^a q^b + 2b_2 \, \pi\indices{^a _c}\pi\indices{^c ^b} \, , \\
 s^a ={}& s u^a +\dfrac{q^a}{T} - \dfrac{1}{2}  (b_1 \, q^b q_b +b_2 \, \pi^{bc}\pi_{bc})\dfrac{u^a}{T} \, , \\
\end{split}
\end{equation}
where $b_1$ and $b_2$ are two transport coefficients. We immediately notice something interesting: the presence of the second order terms ``$\, b_1 \, q^a q^b \,$'' and ``$\, 2b_2 \, \pi\indices{^a _c}\pi\indices{^c ^b} \,$'' in the stress-energy tensor implies that the viscous stress tensor $\pi^{ab}$ is \textit{not} the non-equilibrium part of the total stress tensor. Instead, $\pi^{ab}$ is just an effective field that we use to parameterize such non-equilibrium part. This is in line with what we said before: the physical interpretation of the fields $\varphi^A$ is uniquely and unambiguously fixed only by the constitutive relations.

One of the most useful facts that is highlighted by a field theory approach to hydrodynamics is that we can always redefine the fields $\varphi^A$ without affecting the physics of the system \cite{GavassinoQuasi2022}. 
Thus, let us see what happens if we introduce the ``Israel-Stewart viscous stress'':
\begin{equation}\label{israelfieldredef}
\tilde{\pi}^{ab}:=T_{cd} h^{ac}h^{db}-Ph^{ab}= \pi^{ab}+ b_1 \, q^a q^b + 2b_2 \, \pi\indices{^a _c}\pi\indices{^c ^b} \, .
\end{equation}
If $q^a$ and $\pi^{ab}$ can be treated as small perturbations (i.e. if the fluid is close to local thermodynamic equilibrium \cite{GavassinoFronntiers2021}), we can express \eqref{ConstitutiveGeneric} in terms of $\tilde{\pi}^{ab}$ through a Taylor expansion:
\begin{equation}\label{ConstitutiveGeneric2}
\begin{split}
n^a ={}& n u^a \, , \\
 T^{ab} ={}& \rho u^a u^b +P h^{ab}+q^a u^b+ u^a q^b+\tilde{\pi}^{ab} \, , \\
 s^a ={}& s u^a +\dfrac{q^a}{T} - \dfrac{1}{2}  (b_1 \, q^b q_b +b_2 \, \tilde{\pi}^{bc}\tilde{\pi}_{bc})\dfrac{u^a}{T} + \mathcal{O}(\tilde{\pi}\tilde{\pi}\tilde{\pi})+\mathcal{O}(\tilde{\pi} qq) \, . \\
\end{split}
\end{equation}
If we neglect third-order terms, these are precisely the constitutive relations of the Israel-Stewart theory  in the Eckart frame  (recall that the Eckart frame is defined by the condition $u^a \propto n^a$ \cite{Kovtun2019}). There is, however, an interesting subtlety. The most general entropy current allowed by the Israel-Stewart theory in Eckart frame takes the form \cite{Hishcock1983}
\begin{equation}\label{ISentropy}
 s^a = s u^a +\dfrac{q^a}{T} - \dfrac{1}{2}  (b_0 \, \Pi^2 +b_1 \, q^b q_b +b_2 \, \Pi^{bc}\Pi_{bc})\dfrac{u^a}{T} + a_0 \dfrac{\Pi q^a}{T} + a_1 \dfrac{\Pi^{ab}q_b}{T}  \, .
\end{equation}
As we can see, this entropy current presents five independent transport coefficients $(b_0,b_1,b_2,a_0,a_1)$, while the GENERIC one only two $( b_1,b_2 )$. What about the other three? They are constrained. In fact, if we compare \eqref{ConstitutiveGeneric2} with \eqref{ISentropy}, we see that the entropy currents are the same (to second order) if and only if we set
\begin{equation}
\label{constuo}
b_0 = 3b_2 \, , \spc a_0=a_1=0 \, .
\end{equation}
To obtain the first condition, we invoked the geometric identity $\pi^{ab}\pi_{ab}=\Pi^{ab}\Pi_{ab}+3\Pi^2$, which follows from \eqref{shearbulk} and \eqref{proj}. In conclusion, we can say that  (when $q^a$ and $\pi^{ab}$ are small) the GENERIC constitutive relations are a \textit{particular realization} of the Israel-Stewart constitutive relations, where not all transport coefficients can be chosen independently.

\subsection{Key facts about the field equations}

Clearly, constitutive relations are not enough. To complete the theory, one also needs the field equations (i.e. the equations of motion), to evolve the fields $\varphi^A$. 
In the case of the GENERIC theory, two of such equations are the conservation laws,
\begin{equation}\label{conso}
\begin{split}
& \nabla_a n^a =0 \spc \, \, \, (\text{particle conservation}) \, , \\
& \nabla_a T^{ab} =0 \spc (\text{energy-momentum conservation}) \, , \\
\end{split}
\end{equation} 
and the remaining ones are determined by using the GENERIC formalism \cite{Grmela1997}. Among other things, the GENERIC framework demands the second law of thermodynamics ($\nabla_a s^a \geq 0$) to be \textit{always} respected, by construction. 
Interestingly, the GENERIC entropy production rate has exactly the same form as that of the Israel-Stewart theory (see Appendix \ref{AAA} for the proof), namely \cite{Hishcock1983}
\begin{equation}\label{SecondLaw}
\nabla_a s^a = \dfrac{\Pi^2}{\zeta T} +\dfrac{q^a q_a}{\kappa T^2} + \dfrac{\Pi^{ab}\Pi_{ab}}{2\eta T} \geq 0 \, .
\end{equation}
The (positive) transport coefficients $\zeta$, $\kappa$, and $\eta$ are respectively the bulk viscosity, the heat conductivity, and the shear viscosity. We will discuss the structure of the field equations in more detail in Section \ref{visco?}. For the purpose of answering the question (i) in the Introduction, equations \eqref{conso} and \eqref{SecondLaw} will suffice.

\section{Thermodynamic analysis}
\label{sec:thermo}

In this section, we address question (i): is the theory thermodynamically consistent, stable, and causal? \citet{Stricker2019} have not assessed any of these properties rigorously. They have indeed performed the rest-frame stability analysis, but in relativity the same theory may be stable in one reference frame and unstable in another one \cite{Hiscock_Insatibility_first_order}. Indeed, we have recently shown \cite{GavassinoSuperlum2021} that, if causality is violated (i.e. if information can travel faster than light), then there is some reference frame in which a dissipative theory becomes disastrously unstable. Furthermore, if a theory that obeys the second law of thermodynamics ($\nabla_a s^a \geq 0$) is unstable, then the equilibrium state is not the maximum entropy state \cite{GavassinoLyapunov_2020,GavassinoGibbs2021}, and this leads to thermodynamic inconsistencies. 

Luckily, there is a simple technique that allows us to assess thermodynamic consistency, stability, and causality at the same time \cite{GavassinoCausality2021,GavassinoStabilityCarter2022}. Here, we will apply such technique to the GENERIC theory.

\subsection{Criterion for thermodynamic consistency}
\label{subthermoconst}

Let's start with the problem of thermodynamic consistency. Suppose to bring a liquid into weak contact\footnote{
     By ``weak contact'', we mean that the two systems are allowed to exchange conserved charges, while all the extensive quantities of the total system are the sum of the extensive quantities of the individual parts \cite{GavassinoTermometri}.} 
with a heat-bath ``$H$'', so that the total system ``liquid+bath'' is isolated. Then, the total entropy $S_{\text{tot}}=S+S_H$ cannot decrease \cite{Hishcock1983}, and the total conserved charges $Q_\text{tot}^I=Q^I+Q_H^I$ (e.g., the total number of baryons, the total energy...) are constant \cite{MTW_book,Weinberg_book_1972}. Now, suppose that (for any Cauchy surface we may take) the entropy of the bath can always be expressed via a relation of the form
$S_H=S_H(Q_H^J)$ \cite{GavassinoLorentzInvariance2022}. Then, in the limit in which the heat bath is infinitely larger than the liquid ($|Q^I_H|\gg |Q^I|$), we can expand $S_H$ to first order in the liquid's charges $Q^I$ \cite{huang_book,Termo,GavassinoTermometri}:
\begin{equation}
S_H(Q_H^J)= S_H(Q_\text{tot}^J-Q^J) \approx S_H(Q^J_{\text{tot}})- \dfrac{\partial S_H (Q^J_{\text{tot}})}{\partial Q^I_H} \, Q^I  \, . 
\end{equation}
In fact, all higher-order terms converge to zero, as $Q^I/Q_H^I \rightarrow 0$. Introducing the constant coefficients
\begin{equation}\label{alphastar}
\alpha^\star_I:= - \dfrac{\partial S_H (Q^J_{\text{tot}})}{\partial Q^I_H} \, ,
\end{equation}
the second law of thermodynamics ($\Delta S+\Delta S_H\geq 0$) and the conservation laws ($\Delta Q^I_{\text{tot}}=0$) imply that the function
\begin{equation}
\Phi = S + \alpha^\star_I Q^I 
\end{equation}
is non-decreasing in time\footnote{Note that this is true also if there is no interaction between the liquid and the bath. In fact, because $\alpha^\star_I$ are constants, we have that, in an arbitrary process, $\Delta \Phi= \Delta S +\alpha^\star_I \Delta Q^I$. But if the liquid is isolated, then $\Delta S \geq 0$, and $\Delta Q^I=0$, so that $\Delta \Phi=\Delta S \geq 0$. However, the growth of $\Phi$ is a more general fact, which remains valid also when the liquid is immersed in a larger environment. For example, if the liquid and the bath are separated by a static wall, which is permeable only to the energy ($U$), then the condition $\Delta \Phi \geq 0$ is equivalent to $\Delta (U -T^\star S) \leq 0$, where $T^\star$ is the (constant) temperature of the bath \cite{landau5}.}. On the other hand, the coefficients $\alpha^\star_I$ are constant, as they are functions of the conserved charges $Q^J_{\text{tot}}$ [see equation \eqref{alphastar}].
Thus, as the liquid interacts with the bath, it evolves towards the state that maximizes $\Phi$ for a given value of $\alpha^\star_I$ (specified by the initial conditions of the total system ``liquid+bath''). Such state is what we call ``global thermodynamic equilibrium'' \cite{landau5,Stuekelberg1962,Israel_2009_inbook}. This provides us with a simple criterion for thermodynamic consistency: the functional $\Phi[\varphi^A]= S[\varphi^A] + \alpha^\star_I \, Q^I[\varphi^A] $ must admit a unique maximum (for fixed $\alpha^\star_I$, which may be viewed as Lagrange multipliers \cite{GavassinoGibbs2021}), and such maximum must have all the properties that we ascribe to fluids in thermodynamic equilibrium (e.g. $q^a=\pi^{ab}=0$).

We can rewrite $\Phi[\varphi^A]$ as an integral involving only the physical tensors of the GENERIC theory ($n^a$, $T^{ab}$, and $s^a$). In fact, chosen a spacelike Cauchy 3D-surface $\Sigma$, the functional $\Phi$ evaluated on $\Sigma$ can be expressed as a flux-integral,
\begin{equation}\label{flussusso}
\Phi(\Sigma)= \int_\Sigma \phi^a d\Sigma_a \, ,
\end{equation}
where $d\Sigma_a$ is the volume one-form of $\Sigma$ (we adopt the standard orientation \cite{MTW_book}: $d\Sigma_0 \geq 0$), and
\begin{equation}\label{constructphiA}
\phi^a = s^a +\alpha^\star_I J^{Ia} \, .
\end{equation}
The vector fields $J^{Ia}$ are the conserved currents of the system ($\nabla_a J^{Ia}=0$). By virtue of equation \eqref{conso}, one of these currents is $n^a$ (the corresponding coefficient $\alpha^\star_I$ is simply called ``$\, \alpha^\star \,$''). The remaining currents have the form \cite{Hawking1973}
\begin{equation}\label{conCurrentuzze}
J^{I a}= K^I_b \, T^{ba} \, ,
\end{equation} 
where $K^I_b$ are the Killing vector fields of the spacetime (recall that the spacetime is a fixed background). Introducing the  ``combined'' Killing vector field
$\beta^\star_b =\alpha^\star_I K^I_b$ (with $\alpha^\star_I \neq \alpha^\star$), we can finally express $\phi^a$ as follows:
\begin{equation}\label{gringone}
\phi^a = s^a +\alpha^\star n^a + \beta^\star_b \, T^{ba} \, .
\end{equation}
In Appendix \ref{ThermoKilling}, we calculate explicitly the thermodynamic formula for $\beta^\star_b$ in rotating relativistic stars.

\subsection{Thermodynamic equilibrium}

Our first task is to compute the equilibrium states, as stationary points of the functional $\Phi[\varphi^A]$. The procedure is standard \cite{GavassinoStabilityCarter2022}: we consider a one-parameter family, $\varphi^A(\epsilon)$, of solutions of the field equations, for which $\epsilon=0$ is the equilibrium state. Since the spacetime is treated as  a fixed background, the metric tensor $g_{ab}$ and the Killing vectors $K^I_b$ are held constant (i.e. they do not depend on $\epsilon$). Furthermore, since $\Phi$ attains its maximum for fixed $\alpha^\star_I$, we must keep constant also these. This implies that, in equation \eqref{gringone}, the coefficients $\alpha^\star$ and $\beta^\star_b$ do not depend on $\epsilon$ (we use the superscript ``$\, \star \,$'' to keep track of this distinction \cite{Pathria2011}). Adopting the notation $\dot{f}:=df/d\epsilon$, we need to impose
\begin{equation}\label{staziono}
\dot{\Phi}(\epsilon=0)=0 \, ,
\end{equation}
for any choice of one-parameter family (defined as above), and for any choice of spacelike Cauchy 3D-surface $\Sigma$. Invoking equations \eqref{ConstitutiveGeneric}, \eqref{flussusso}, and \eqref{gringone}, and introducing the compact notation  $b^a_i := b_i u^a/2T$, we can express respectively $\Phi$, $\dot{\Phi}$, and $\ddot{\Phi}$ as fluxes of the currents reported below (for any $\epsilon$): 
\begin{fleqn}
\begin{equation}
\begin{split}
\phi^a ={}& \big[s +\alpha^\star n + \beta^\star_b u^b (\rho+P) \big]u^a +P \beta^{\star a}  
+ \beta^\star_b q^b u^a + \beta^\star_b \pi^{ba} + ( T^{-1}  + \beta^\star_b u^b)q^a 
\\
& - b_1^a q^b q_b  + b_1 q^a q^b \beta^\star_b - b_2^a \pi^{bc}\pi_{bc} +2b_2 \pi^a_c\pi^{cb}\beta^\star_b \, ,  
\end{split} 
\end{equation}
\begin{equation}
\begin{split}
\dot{\phi}^a ={ } & \big[ \dot{s}+\alpha^\star \dot{n} + \beta_b^\star u^b (\dot{\rho}+\dot{P})  + \beta_b^\star \dot{u}^b (\rho+P) \big]u^a+ \big[ s+\alpha^\star n + \beta_b^\star u^b (\rho+P) \big]\dot{u}^a +\dot{P}\beta^{\star a} 
\\
&  + \beta^\star_b \dot{q}^b u^a+\beta^\star_b q^b \dot{u}^a + \beta^\star_b \dot{\pi}^{ba} +(T^{-1}  + \beta^\star_b u^b)\dot{q}^a+( -T^{-2}\dot{T}  + \beta^\star_b \dot{u}^b)q^a 
\\
& - \dot{b}_1^a q^b q_b-2b_1^a q^b \dot{q}_b  + \dot{b}_1 q^a q^b \beta^\star_b+b_1 \dot{q}^a q^b \beta^\star_b +b_1 q^a \dot{q}^b \beta^\star_b 
\\
& - \dot{b}_2^a \pi^{bc} \pi_{bc} -2b_2^a \pi^{bc}\dot{\pi}_{bc} +2\dot{b}_2 \pi^a_c\pi^{cb}\beta^\star_b + 2b_2 \dot{\pi}^a_c\pi^{cb}\beta^\star_b +2b_2 \pi^a_c\dot{\pi}^{cb}\beta^\star_b \, ,  
\end{split}
\end{equation}
\begin{equation}
\begin{split}
\ddot{\phi}^a ={ } & \big[ \ddot{s}+\alpha^\star \ddot{n} + \beta_b^\star u^b (\ddot{\rho}+\ddot{P})  + \beta_b^\star \ddot{u}^b (\rho+P) +2\beta_b^\star \dot{u}^b (\dot{\rho}+\dot{P}) \big]u^a+ \big[ s+\alpha^\star n + \beta_b^\star u^b (\rho+P) \big]\ddot{u}^a 
\\ 
& +2 \big[ \dot{s}+\alpha^\star \dot{n} + \beta_b^\star u^b (\dot{\rho}+\dot{P})  + \beta_b^\star \dot{u}^b (\rho+P) \big]\dot{u}^a+\ddot{P}\beta^{\star a} + \beta^\star_b \ddot{q}^b u^a+2 \beta^\star_b \dot{q}^b \dot{u}^a+\beta^\star_b q^b \ddot{u}^a + \beta^\star_b \ddot{\pi}^{ba} 
\\
& + ( T^{-1}  + \beta^\star_b u^b)\ddot{q}^a+2( -T^{-2}\dot{T}  + \beta^\star_b \dot{u}^b)\dot{q}^a+( 2 T^{-3}\dot{T}^2 -T^{-2}\ddot{T}  + \beta^\star_b \ddot{u}^b)q^a 
\\
& - \ddot{b}_1^a q^b q_b-2b_1^a q^b \ddot{q}_b - 4\dot{b}_1^a q^b \dot{q}_b - 2b_1^a \dot{q}^b \dot{q}_b   - \ddot{b}_2^a \pi^{bc} \pi_{bc} -2b_2^a \pi^{bc}\ddot{\pi}_{bc}-4\dot{b}_2^a \pi^{bc}\dot{\pi}_{bc} -2b_2^a \dot{\pi}^{bc}\dot{\pi}_{bc} 
\\
& + \ddot{b}_1 q^a q^b \beta^\star_b+b_1 \ddot{q}^a q^b \beta^\star_b +b_1 q^a \ddot{q}^b \beta^\star_b +2 b_1 \dot{q}^a \dot{q}^b \beta^\star_b + 2\dot{b}_1 \dot{q}^a q^b \beta^\star_b + 2\dot{b}_1 q^a \dot{q}^b \beta^\star_b
\\
& +2\ddot{b}_2 \pi^a_c\pi^{cb}\beta^\star_b + 2b_2 \ddot{\pi}^a_c\pi^{cb}\beta^\star_b +2b_2 \pi^a_c\ddot{\pi}^{cb}\beta^\star_b +4b_2 \dot{\pi}^a_c\dot{\pi}^{cb}\beta^\star_b +4\dot{b}_2 \dot{\pi}^a_c\pi^{cb}\beta^\star_b +4\dot{b}_2 \pi^a_c\dot{\pi}^{cb}\beta^\star_b \, .  
\end{split}
\end{equation}
\end{fleqn}
Since equation \eqref{staziono} must be true for any choice of one-parameter family $\varphi^A(\epsilon)$, and for any spacelike Cauchy 3D-surface $\Sigma$, we need to set $\dot{\phi}^a(\epsilon=0)=0$, for any allowed variation $\{\dot{\rho},\dot{n},\dot{u}^a,\dot{q}^a,\dot{\pi}^{ab} \}$. Recalling the structural identities in \eqref{algconstraints} and \eqref{thermo}, it can be easily verified  that the above happens if and only if the equilibrium state ($\epsilon=0$) satisfies the identities below\footnote{The reader can verify explicitly that the equilibrium states \eqref{equilibrioconduzio} are indeed solutions of the field equations of the GENERIC theory.}:
\begin{equation}\label{equilibrioconduzio}
\mu/T = \alpha^\star \spc u_b/T = \beta^\star_b \spc q^a=\pi^{ab}=0 \, .
\end{equation}
Recalling that $\alpha^\star$ is a constant, and $\beta^\star_b$ is a Killing vector field, we can conclude that, at equilibrium, $\nabla_a(\mu/T)=0$ and $\nabla_{a}(u_b/T)+\nabla_b (u_a/T)=0$. Both these equilibrium conditions are consistent with relativistic thermodynamics \cite{Israel_Stewart_1979} and statistical mechanics \cite{BecattiniBeta2016}. Furthermore, we see from equation \eqref{SecondLaw} that, at equilibrium, the fluid's evolution is reversible (namely $\nabla_a s^a =0$), as it should be. Finally, we can use \eqref{equilibrioconduzio} to show that
\begin{equation}\label{Hawk}
\Phi(\epsilon=0)= \int_\Sigma P \beta^\star_a d\Sigma^a \, ,
\end{equation}
which is consistent with the corresponding formula of \citet{GibbonsHawking1977}. 
In conclusion, the equilibrium properties of the GENERIC theory are fully consistent with thermodynamics. 
 
\subsection{Thermodynamic inequalities}

In the previous subsection, we have identified the equilibrium state by requiring that it makes the functional $\Phi$ stationary: $\dot{\Phi}(\epsilon=0)=0$. However, we still need to make sure that the state $\epsilon=0$ is a genuine maximum of $\Phi$. In practice, this amounts to requiring that the functional $E=\Phi(0)-\Phi(\epsilon)$ is non-negative definite (for any $\epsilon$). In the limit of small $\epsilon$, this functional can be expressed as the flux of an associated current,
\begin{equation}
E^a = \phi^a(0)-\phi^a(\epsilon) = -\dfrac{1}{2}\ddot{\phi}^a(0) \epsilon^2 + \mathcal{O}(\epsilon^3) \, ,
\end{equation} 
which is known as the ``information current'' \cite{GavassinoCausality2021}. As we can see, to compute $E^a$, we only need to evaluate $\ddot{\phi}^a$ at equilibrium. Luckily, as a result of the constraints \eqref{algconstraints}, and of their derivatives (all evaluated at equilibrium),
\begin{equation}
\begin{split}
& u_a \dot{u}^a=u_a \dot{q}^a= u_a \dot{\pi}^{ab}=\dot{\pi}^{[ab]}=0 \, ,\\
& u^a \ddot{u}_a +\dot{u}^a \dot{u}_a =  u_a \ddot{q}^a+2\dot{u}_a \dot{q}^a=u_a \ddot{\pi}^{ab} + 2\dot{u}_a \dot{\pi}^{ab}=\ddot{\pi}^{[ab]}=0 \, , \\
\end{split}
\end{equation}
the formula for $\ddot{\phi}^a(0)$ is much simpler than the general formula of $\ddot{\phi}^a$ for arbitrary $\epsilon$:
\begin{equation}
\begin{split}
T\ddot{\phi}^a(0) ={ } & -\big[ \dot{T} \dot{s}+\dot{\mu} \dot{n}   + \dot{u}_b \dot{u}^b (\rho+P)\big]u^a 
-2 \dot{P}\dot{u}^a -2\dot{u}_b \dot{q}^b u^a- 2\dot{u}_b \dot{\pi}^{ba} \\
& -2T^{-1}\dot{T} \dot{q}^a  - b_1 u^a \dot{q}^b \dot{q}_b    -b_2 u^a \dot{\pi}^{bc}\dot{\pi}_{bc}  \, . \\
\end{split}
\end{equation}
With the aid of \eqref{dPippuz} and \eqref{dhuzz}, the first two terms in the square brackets can be rewritten as follows \cite{GavassinoCausality2021}:
\begin{equation}
\dot{T}\dot{s}+\dot{\mu}\dot{n}= \dfrac{nT}{c_p} \dot{\mathfrak{s}}^2 + \dfrac{\dot{P}^2}{c_s^2(\rho+P)} \, .
\end{equation}
Thus, introducing the standard notation $\delta f:=f(\epsilon)-f(0)= \dot{f}(0) \, \epsilon +\mathcal{O}(\epsilon^2)$, the information current takes the form
\begin{equation}
\begin{split}
TE^a ={ } & \dfrac{1}{2} \bigg[  \dfrac{nT}{c_p} (\delta\mathfrak{s})^2 + \dfrac{(\delta P)^2}{c_s^2(\rho+P)}   +  (\rho+P) \delta u_b \delta u^b+2\delta u_b \delta q^b + b_1  \delta q^b \delta q_b    +b_2  \delta \pi^{bc}\delta \pi_{bc} \bigg]u^a \\
& +\dfrac{\delta T \delta q^a}{T} + \delta P\delta u ^a + \delta u_b \delta \pi^{ba} +\mathcal{O}(\epsilon^3)   \, . \\
\end{split}
\end{equation} 
This is \textit{exactly} the information current of the Israel-Stewart theory \cite{Hishcock1983,GavassinoGibbs2021,Almaalol2022,GavassinoUniversality2023}, subject to the constraints \eqref{constuo}. If we think about it, this is no surprise: the above formula for $E^a$ depends on the constitutive relations up to second order in deviations from equilibrium. But the differences between the GENERIC constitutive relations and Israel-Stewart ones appear only at third order [see equation \eqref{ConstitutiveGeneric2}].

For the theory to be thermodynamically consistent, the functional $E$, defined as the flux of the current $E^a$, must be strictly positive for any non-vanishing perturbation, and for any Cauchy surface $\Sigma$. This produces a set of thermodynamic inequalities which, for the Israel-Stewart theory, have already been computed by \citet{Hishcock1983}. Since our current $E^a$ is identical to theirs, we can just ``copy'' their formulas, implementing the constraints \eqref{constuo} where needed. The result are 8 inequalities of the form $\Omega_i >0$, $i=1,...,8$ \cite{Hishcock1983}, with
\begin{equation}\label{Omeguna}
\begin{matrix*}[l]
& \Omega_1 = \dfrac{1}{c_s^2 (\rho+P)} \, , \quad & \Omega_2 = \dfrac{nT}{c_p} \, , \quad & \Omega_3 = (\rho +P )(1-c_s^2) - \dfrac{1}{b_2} - \dfrac{K^2}{\Omega_6} \, , \quad & \Omega_4 = \rho+P - \dfrac{2b_2 +b_1}{2b_1 b_2} \, , \\
& \Omega_5 = 3b_2 \, , \quad & \Omega_6 = b_1 - \dfrac{1}{nTc_v} \, , \quad & \Omega_7 = b_1 \, , \quad & \Omega_8 = b_2 \, . \\
\end{matrix*}
\end{equation}
In the above expression for $\Omega_6$, $c_v$ is the specific heat at constant volume. In $\Omega_3$, the expression $K$ is defined as
\begin{equation}
K = 1 - \dfrac{(\rho+P)k_p c_s^2}{nc_p} \, .
\end{equation}
The thermodynamic constraints that arise from the positivity of all the $\Omega_i$ are quite complicated. However, there are some notable inequalities which are worth discussing. From the positivity of $\Omega_7$ and $\Omega_8$ we infer that $b_1$ and $b_2$ are positive. This implies that the physical entropy density $-u_a s^a$ is smaller than its equilibrium part $s$, as one would expect \cite{Israel_Stewart_1979}. Furthermore, the positivity of $\Omega_4$ guarantees that $\rho+P>0$ (positive inertia \cite{MTW_book}), while the positivity of $\Omega_1$ enforces the inequality $c_s^2>0$ (stability against adiabatic compression). The positivity of $\Omega_3$ implies that the adiabatic sound-speed cannot exceed the speed of light. The positivity of $\Omega_2$ implies $c_p>0$, which is a thermodynamic inequality valid for any extensive system \cite{landau5}. Indeed, \citet{Hishcock1983} have shown that all the ``textbook'' thermodynamic inequalities (e.g. positive specific heats and compressibilities) follow directly from the positivity of all the $\Omega_i$. Also the (relativistic) Schwarzschild criterion for stability of the equilibrium against convection is automatically respected \cite{Hishcock1983}. Finally, from the positivity of $\Omega_4$ we recover the inequality $b_1 > (\rho+P)^{-1}$, see \cite{GavassinoLyapunov_2020}.

\subsection{Stability and causality}

In a relativistic setting, for theories that obey the second law $\nabla_a s^a \geq 0$, thermodynamic consistency implies (covariant) stability \cite{GavassinoGibbs2021} and causality \cite{GavassinoCausality2021} close to equilibrium. For what concerns stability, the proof is rather easy. Consider again the functional $E=\Phi(0)-\Phi(\epsilon)$. The first term on the right-hand side, $\Phi(0)$, is the quantity $\Phi$ evaluated at equilibrium, and it is a constant, while the second term, $\Phi(\epsilon)$, depends on the hypersurface $\Sigma$. However, from the discussion in subsection \ref{subthermoconst}, we know that $\Phi(\epsilon)$ is non-decreasing in time. This implies that $E$ can only decrease (or stay constant). On the other hand, by thermodynamic consistency, $E$ is a positive definite square-integral norm of the perturbation fields $\delta \varphi^A$ (at least for small $\epsilon$). Thus, the perturbations about the equilibrium state are stable in the sense that they must evolve keeping the norm $E(\Sigma)$ bounded below by 0 and above by its initial value \cite{Hishcock1983}. 

The proof discussed above is a manifestation of Lyapunov's stability theorem (or ``Lyapunov's direct method'' \cite{lasalle1961stability}): if the function $\Phi$ is non-decreasing in time, and it has an isolated maximum at equilibrium, then the equilibrium state is Lyapunov-stable. Indeed, \citet{Grmela1997} themselves had invoked this same principle to prove the stability of the equilibrium state within the GENERIC formalism\footnote{
    The function ``$\Phi$'' introduced by \citet{Grmela1997} in their equation (17) is actually $-\Phi$, according to our notation. Note also that our equation \eqref{Hawk} is the General-Relativistic analogue of equation (28) of \cite{Grmela1997}.
} (see \cite{Grmela1997}, Section B.5, property 3). Here, we have just shown how to apply this same idea to the GENERIC hydrodynamic model of \citet{Stricker2019}. The key distinction with respect to \cite{Grmela1997} is that, here, we define $\Phi$ in terms of an arbitrary spacelike Cauchy 3D-surface $\Sigma$, which may correspond to the $ t=0 $ hypersurface of an arbitrary observer. In this way, we are sure that the fluid is stable in \emph{all} reference frames (i.e. it is \textit{covariantly} stable \cite{GavassinoSuperlum2021,GavassinoBounds2023}). Furthermore, since the metric tensor $g_{ab}$ and the (timelike) Killing vector field $\beta^\star_b$ are arbitrary, we have automatically proved that also rotating equilibria are stable, as well as equilibria in a background gravitational field. 

The connection between thermodynamic consistency and causality \cite{GavassinoCausality2021}, on the other side, comes from the relativity of simultaneity: causality is necessary for the laws of thermodynamics to be Lorentz invariant. This was noticed for the first time in \cite{GavassinoLyapunov_2020}, and is discussed extensively in \cite{GavassinoSuperlum2021}.

In conclusion, we can safely say that the answer to question (i) in the Introduction is ``yes'': the theory in \cite{Stricker2019} is thermodynamically consistent, stable, and causal (at least close to thermodynamic equilibrium), provided that all the eight coefficients $\Omega_i$ listed in equation \eqref{Omeguna} are positive.

\section{Viscoelasticity?}
\label{visco?}

Now we can move to question (ii): does the GENERIC theory exhibit a viscoelastic behavior, which would make it well-suited for modeling relativistic liquids? Again, we will limit our analysis to small deviations from equilibrium. For simplicity, we assume (only in this section) that the spacetime is Minkowski, and the background equilibrium state is homogeneous.

\subsection{Variational principle}

In \cite{GavassinoQuasi2022}, we have shown that, if a thermodynamically-consistent theory is of Geroch-Lindblom type \cite{Geroch_Lindblom_1991_causal} (i.e. if it admits a symmetric-hyperbolic formulation), then it is possible to ``guess'' its linearised field equations, by means of a variational principle:
\begin{equation}\label{trentotto}
T \, \nabla_a \dfrac{\partial E^a}{\partial (\delta \varphi^A)} = -\dfrac{T}{2} \dfrac{\partial \sigma}{\partial (\delta \varphi^A)} \, ,
\end{equation}
where $E^a$ is the information current, and $\sigma:=\nabla_a s^a$ is the entropy production rate, both expressed up to second order in $\delta \varphi^A$. We still don't know whether the GENERIC theory is of Geroch-Lindblom type, but let's just apply this variational principle, and see what happens. Following the procedure outlined in \cite{GavassinoQuasi2022}, we work in the background's rest frame, where the second-order expressions for $E^a$ and $\sigma$ become
\begin{equation}
\begin{split}
E^0 =  &  \dfrac{1}{2T} \bigg[  \dfrac{nT}{c_p} (\delta\mathfrak{s})^2 + \dfrac{(\delta P)^2}{c_s^2(\rho+P)}   +  (\rho+P) \delta u_j \delta u^j+2\delta u_j \delta q^j +3b_2 (\delta \Pi)^2+ b_1  \delta q^j \delta q_j    +b_2  \delta \Pi^{jk}\delta \Pi_{jk} \bigg] \, , 
\\
E^j =  &  \dfrac{1}{T} \bigg[\dfrac{\delta T \delta q^j}{T} + \delta P\delta u ^j+\delta \Pi \delta u ^j +  \delta \Pi^{jk} \delta u_k \bigg] \, , 
\\
\sigma  =  & \dfrac{(\delta\Pi)^2}{\zeta T} +\dfrac{\delta q^j \delta q_j}{\kappa T^2} + \dfrac{\delta \Pi^{jk}\delta\Pi_{jk}}{2\eta T} \, . 
\end{split}
\end{equation}
Then, we derive the field equations using \eqref{trentotto}. However, instead of performing the variations in the variables $\delta\varphi^A = \{\delta\rho, \delta n, \delta u^j , \delta q^j , \delta \pi^{jk} \}$, it is more convenient to take as our degrees of freedom the variables
\begin{equation}
\delta \tilde{\varphi}^B = \{ \delta\mathfrak{s}, \delta P, \delta u^j ,\delta \Pi, \delta q^j , \delta \Pi^{jk} \} \, .
\end{equation}
We are allowed to do this, because the variational principle \eqref{trentotto} is invariant under field redefinitions. The guessed field equations, computed using \eqref{trentotto}, are reported below. 
Following \citep{GavassinoQuasi2022}, the field with respect to which we perform the variation is written inside a box in front of the corresponding equation:
\begin{flalign}\label{settatntuno}
\boxed{\mathfrak{s}} \quad \dfrac{nT}{c_p} \partial_t \delta \mathfrak{s} + \dfrac{\partial_j \delta q^j}{c_p}=0 &&
\end{flalign}
\begin{flalign}\label{settantadue}
\boxed{P} \quad \dfrac{\partial_t \delta P}{(\rho+P)c_s^2} + \partial_j \delta u^j + \dfrac{\kappa_p \partial_j \delta q^j}{nc_p}=0 &&
\end{flalign}
\begin{flalign}\label{settantatre}
\boxed{u^k} \quad \partial_t[ (\rho + P)  \delta u_k +  \delta q_k] + \partial_k (\delta P + \delta \Pi) + \partial_j \delta \Pi^j_k = 0  &&
\end{flalign}
\begin{flalign}\label{settantquattro}
\boxed{\Pi} \quad 3b_2 \, \partial_t \delta \Pi  + \partial_j \delta u^j  = -\dfrac{\delta \Pi}{\zeta} &&
\end{flalign}
\begin{flalign}\label{settantcinque}
\boxed{q^k} \quad  \partial_t(b_1  \delta q_k + \delta u_k) + \dfrac{\partial_k \delta T}{T}  =  -\dfrac{\delta q_k}{\kappa T} &&
\end{flalign}
\begin{flalign}\label{settantasei}
\boxed{\Pi^{kl}} \quad b_2 \, \partial_t \delta \Pi_{kl} + \braket{\partial_k \delta u_l}= -\dfrac{\delta \Pi_{kl}}{2\eta}  &&
\end{flalign}
Here, $\braket{A_{kl}}$ denotes the symmetric traceless part of $A_{kl}$. The equations above are the linearised field equations of the Israel-Stewart theory \cite{Hishcock1983}, under the constraint \eqref{constuo}. Indeed, it could not be otherwise, since both the information current $E^a$ and the entropy production rate $\sigma$ of the GENERIC theory are indistinguishable from those of the Israel-Stewart theory, whose field equations (in the linear regime) are determined by the variational procedure \eqref{trentotto} \cite{GavassinoQuasi2022}.

A less obvious fact is that equations \eqref{settatntuno}-\eqref{settantasei} are also the linearised field equations of the GENERIC theory! In fact, the first three equations are the linear limit of \eqref{conso} and \eqref{SecondLaw}. This can be easily checked by linearising the constitutive relations \eqref{ConstitutiveGeneric}. The remaining three equations are the linear limit of the field equations for the structural variables ``$\omega_a$'' and ``$\alpha_{ab}$'' considered by \citet{Stricker2019}. We verify this explicitly in Appendix \ref{equiWuz}, with the aid of the notational correspondence laid down in Appendix \ref{AAA}. The implication is simple: in the linear regime, the GENERIC theory is a particular realization of the Israel-Stewart theory, with the additional assumption that $b_0 = 3b_2$, and $a_0=a_1=0$.

\subsection{Consequences of the equivalence}

The fact that (in the linear regime) the GENERIC theory is a particular example of Israel-Stewart theory has several interesting implications. Below, we list some of them:
\begin{itemize}
\item In their study of the dispersion relations of the linearised GENERIC theory, \citet{Stricker2019} have found several modes that have been previously identified by \citet{Pu2010}, who were studying the Israel-Stewart theory. The reason is now clear: the dispersion relations of the two linearised theories are the same.
\item Equations \eqref{settatntuno}-\eqref{settantasei} constitute a symmetric-hyperbolic system of equations. This guarantees that, in the linear regime, the initial-value problem of the GENERIC theory is well-posed. This just means that the solution to the field equations exists, is unique, and depends continuously on the initial data \cite{Causality_bulk}.
\item We have further corroborated the idea that the Israel-Stewart theory represents a universality class of hydrodynamic theories. In fact, the linearised Israel-Stewart theory (or one of its particular realizations) is recovered whenever we take the linear limit of a large class of non-linear models, including divergence-type theories \cite{Liu1986,GerochLindblom1990}, Carter's theory \cite{PriouCOMPAR1991,GavassinoKhalatnikov2022}, relativistic fluids with chemical or nuclear reactions \cite{BulkGavassino,Camelio2022,CamelioSecondo2022}, and now also the GENERIC theory.
\item As we can see from equations \eqref{settantquattro} and \eqref{settantasei}, the linearised GENERIC theory is just Maxwell's model for viscoelasticity. For low-frequency perturbations, it describes a viscous system, while, for high-frequency perturbations, it describes an elastic system (see Appendix \ref{kjhA}). This supports the idea of \citet{Stricker2019} that the GENERIC theory may be well suited for describing liquids. However, in this regard, the GENERIC theory is not superior to Israel-Stewart (in the linear regime). 
\end{itemize}
In conclusion, the answer to question (ii) of Sec. \ref{intro} is again ``yes'': the expected viscous and elastic behaviors are indeed recovered, in the appropriate limits.

\section{Possible applications to relativistic fluids}
\label{sec:new}

We still have to answer question (iii) in the Introduction: is the GENERIC theory somehow better suited to liquid modeling than Israel-Stewart? One possible way of addressing this question may be to see whether the theory presents, in the non-linear regime, some far-from-equilibrium properties that are characteristic of liquids. 
However, far-from-equilibrium effects are typically system-specific (different substances may behave in very different ways). Hence, we will stick to the linear theory, where more general conclusions can be drawn.
Indeed, any heuristic model for the full non-linear dynamics should boil down to a linearised theory when we are close enough to equilibrium, and while the admissibility of the theory in the linear regime does not tell us much about the admissibility of the fully nonlinear one, the non-admissibility of the theory in the linear regime automatically tells us that also the full one should be disregarded.
Therefore, we will focus here on what singles out the GENERIC theory from the broader family of Israel-Stewart-like models: equation \eqref{constuo}, and in particular the constraint $b_0=3b_2$. 
Is this identity a distinctive feature of liquids close to equilibrium, or is it just a shortcoming of the specific construction of the GENERIC theory? 

It is useful to note that the condition $b_0=3b_2$ can be equivalently expressed in two other ways. One way is in terms of the relaxation times:
\begin{equation}
\label{ITAU}
\dfrac{\tau_\zeta}{\tau_\eta} = \dfrac{3\zeta}{2\eta} \, ,
\end{equation}
where $\tau_\zeta$ and $\tau_\eta$ are the bulk and shear relaxation time. To derive \eqref{ITAU}, one only needs to recast equations \eqref{settantquattro} and \eqref{settantasei} in the standard Cattaneo's \cite{cattaneo1958} form: $\tau \partial_t f+f \propto -\nabla z$. The other way, instead, is to compare the high-frequency limit of \eqref{settantquattro} and \eqref{settantasei} with the corresponding formulas provided by the theory of elasticity \cite{landau7}: $\Pi=-G_\zeta \varepsilon\indices{^j _j}$, where $G_\zeta$ is the bulk modulus, and $\Pi_{kl}=-2G_\eta \braket{\varepsilon_{kl}}$, where $G_\eta$ is the shear modulus. This produces the following constraint:
\begin{equation}
\lambda := G_\zeta - \dfrac{2}{3}G_\eta =0 \, .
\end{equation}
The quantity $\lambda$ is known as ``Lam\'{e}'s first parameter'' \cite{landau7}, and according to the GENERIC theory it vanishes. 
In the following, we discuss if this is the case for some interesting systems that are often studied by means of relativistic hydrodynamic models.

\subsection{Ideal gases}

We are mainly interested in liquids. However, to ``warm up'', let us see if the constraint $b_0=3b_2$ is valid for ideal gases. \citet{Israel_Stewart_1979} have computed the quantities $b_0$ and $b_2$ using Grad's 14-moment approximation, for a non-degenerate gas of particles with mass $m$. In the non-relativistic limit, they obtained
\begin{equation}
b_0 = \dfrac{6m^2}{5 T^2 P} \, , \spc b_2 = \dfrac{1}{2P} \, , 
\end{equation}
while, in the ultra-relativistic limit, they found
\begin{equation}
b_0 = \dfrac{216 \, T^4}{ m^4 P} \, , \spc  b_2 = \dfrac{3}{4P} \, . 
\end{equation}
Clearly, $b_0$ does not coincide with $3b_2$: since the constraint \eqref{constuo} is not fulfilled, the GENERIC theory cannot be applied to ideal gases.

\subsection{Neutron-star matter}

When the astrophysical process under consideration does not drive matter too far from equilibrium (e.g. by pushing the $\beta$-reaction affinity above $T$, see e.g. \citep{Schmitt2018ASSL}), the fluid interior of a neutron star can be modelled as a substance belonging to the Israel-Stewart class \citep{Camelio2022,CamelioSecondo2022}, or a natural extension of it that accounts for the possible presence of superfluid neutrons \citep{GavassinoKhalatnikov2022}.

Differently from the ideal gas case discussed above, for strongly interacting nuclear matter there are no analytical results for $b_2$. However, it is unlikely that the GENERIC relation \eqref{ITAU} will be fulfilled in neutron stars. This is due to the variety of possible processes giving rise to bulk and shear viscosity, so that different channels can be more effective (or suppressed) according to the local thermodynamic state of matter (see \citep{Schmitt2018ASSL} for a review on transport in neutron stars). The GENERIC model would be a viable option for neutron stars only in the event of an accidental tuning between the many (practically unrelated) physical processes that give rise to bulk and shear viscosity separately. 

In a neutron star, bulk viscosity mostly comes from out-of-equilibrium nuclear reactions  and, where superfluid contributions are relevant in low temperature range, reactions involving excitations of the neutron superfluid can also contribute. On the other hand, shear viscosity arises from a completely different set of microscopic processes.
For cores composed of strongly degenerate neutrons, protons, electrons, and muons, the main contribution to the shear viscosity is leptonic (the most mobile particles) with corrections due to neutrons (the most abundant species),  see e.g. \citep{Shternin_core_08}. 
Therefore, while bulk viscosity is mostly due to nuclear reactions controlled by the weak sector, the shear viscosity is mainly given by scattering of electrons and muons between themselves and with protons, via the electromagnetic force. 
On top of this, the neutron contribution to shear viscosity is defined by neutron-neutron and neutron-proton collisions mediated by the nuclear strong interaction. 

Finally, at very low temperatures, superfluid phonon modes can dominate the thermal corrections to the thermodynamic and hydrodynamic properties of the superfluid core: the superfluid phonon contribution could be relevant for the determination of both the shear and bulk viscosities, as well as the thermal conductivity \citep{EscobedoManuel2009,Manuel2021Univ}.
However, even disregarding completely some features of superfluid matter (i.e., the possible presence of extra currents \citep{Termo,GavassinoKhalatnikov2022}) and restricting ourselves to the physical regime where the standard Israel-Stewart theory is applicable \citep{BulkGavassino,Camelio2022}, it seems unlikely that the constraint \eqref{constuo} could be physically fulfilled in neutron star interiors. 


\subsection{Quark-gluon plasma}

\citet{Stricker2019} have suggested that the GENERIC theory may be well-suited for modelling the Quark-Gluon Plasma (QGP). Unfortunately, holography-based studies \cite{KovtunHolography2005,Noronha2011,Rougemont2021} have revealed that the QGP probably does not belong to the Israel-Stewart ``universality class'' (even close to equilibrium). Instead, the QGP seems to exhibit a rather different kind of non-equilibrium dynamics \cite{Heller2014}, which constitutes to a universality class of its own (alternative to Maxwell's model), where, instead of postulating a relaxation-type field equation like \eqref{Israel1}, one postulates the existence of a conjugate degree of freedom $\Lambda_{xy}$, which is coupled to the stress $\Pi_{xy}$ through a skew-symmetric interaction \cite{GavassinoQuasi2022}:
\begin{equation}
\begin{bmatrix}
 b_2 \, \partial_t \Pi_{xy} + \partial_{(x} u_{y)} \\
  b_2 \,  \partial_t\Lambda_{xy} \\
\end{bmatrix}
= 
\begin{bmatrix}
 0 & & \mathcal{A} \\
 -\mathcal{A} & & -\xi_3 \\
\end{bmatrix}
\begin{bmatrix}
 \Pi_{xy} \\
 \Lambda_{xy} \\
\end{bmatrix} \, .
\end{equation}
The Israel-Stewart formalism cannot reproduce this kind of dynamics \cite{Denicol_Relaxation_2011}. In fact, in the absence of spatial gradients, the Maxwell model predicts that $\Pi_{xy}$ will relax exponentially to zero, over a timescale $\tau_\eta$. By contrast, the holographic equation above predicts that $\Pi_{xy}$, besides relaxing towards its Navier-Stokes value, will also oscillate, with frequency
\begin{equation}
\omega_{\text{QNM}}=\dfrac{\sqrt{4\mathcal{A}^2 - \xi_3^2}}{2b_2}  \, ,
\end{equation}
which is real, because $2\mathcal{A}> \xi_3$ (in the QGP). Since the GENERIC theory, as formulated in \citep{Stricker2019}, belongs to the Israel-Stewart universality class, it is not suitable for modeling the dynamics of the QGP.

\section{Conclusions}

We have shown that the GENERIC hydrodynamic theory developed in \citep{Stricker2019} for relativistic liquids belongs (in the linear regime) to the universality class defined by the Israel-Stewart theory. In fact, despite the Israel-Stewart theory was originally devised to describe dilute relativistic gases, it is a valid phenomenological description also of liquid substances, as it can reproduce the high-frequency ``elastic'' behavior of liquids \citep{Frenkel_book}, and is the relativistic extension of the Maxwell model for viscoelasticity \citep{Nettleton1959},  see Appendix \ref{kjhA}. Furthermore, we have also presented a dictionary to pass from the GENERIC theory of \cite{Stricker2019} to EIT. 
This should help build further bridges between the GENERIC formalism and other frameworks used in relativistic hydrodynamics, similarly to what has been already done for Carter's theory and EIT~\citep{GavassinoFronntiers2021,GavassinoKhalatnikov2022}.

More precisely, we have shown that, in the linear regime, the field equations of the GENERIC theory form a symmetric-hyperbolic system, and they constitute a particular realization of the Israel-Stewart theory (which is also known to be symmetric-hyperbolic \cite{Hishcock1983}), where the bulk and shear relaxation times are related by formula \eqref{ITAU}. 

In conclusion, did GENERIC pass the tests outlined in (i-ii-iii)? The GENERIC \textit{formalism}, regarded as a methodology for constructing hydrodynamic theories, passed the tests very well. In fact, it leads to theories that are causal, stable, symmetric-hyperbolic, and thermodynamically consistent (in the linear regime). 
On the other hand, the specific GENERIC hydrodynamic theory in \citep{Stricker2019} makes predictions - the identities \eqref{constuo} or \eqref{ITAU} - that are inconsistent with the microphysics of many substances (even close to equilibrium).
Indeed, as discussed in Sec. \ref{sec:new}, we could not find any type of relativistic fluid whose near-equilibrium behaviour could be described by this specific GENERIC theory. 
Despite this apparent failure of the theory, this does not mean that the formalism and the ideas implemented in its construction have no value.  
On the contrary, considering that GENERIC is primarily a formalism \cite{Grmela1997,OttingerReview2018} rather than a specific theory, we believe that, given the present analysis, it is now possible to fix the specific theory of \citet{Stricker2019} to make it well-suited for modelling both liquids and gases.  
We also expect that, by expanding the number of algebraic degrees of freedom (above the 14 considered in \cite{Stricker2019}) one may also give rise to a broader class of GENERIC hydrodynamic models which do not necessarily reduce to  Israel-Stewart close to equilibrium. Indeed, application of the Onsager-Casimir principle to relativistic hydrodynamics has recently revealed \cite{GavassinoCasimir2022} that the GENERIC formalism always reduces, in the linear regime, to a subset of the general symmetric-hyperbolic quasihydrodynamic framework derived in \cite{GavassinoQuasi2022}.  

\section*{Acknowledgements}

L.G. is partially supported by a Vanderbilt’s Seeding Success Grant.
M.A. acknowledges partial support from PHAROS, COST Action CA16214. L.G. would like to thank Jean-Francois Paquet and Weiyao Ke for stimulating discussions.

\appendix

\section{Israel-Stewart is viscoelastic}
\label{kjhA}

We briefly discuss the non-Newtonian properties of the Israel-Stewart model.
To do so, consider the following linear system of partial differential equations:
\begin{equation}\label{camusa}
    \begin{split}
 & (\rho+P)\partial_t  u_y +\partial_x  \Pi_{xy}=0 \, , \\
 & \tau_\eta \partial_t \Pi_{xy}+\Pi_{xy}=-\eta \partial_x u_y \, , \\
    \end{split}
\end{equation}
where $\rho$, $P$, $\tau_\eta$, and $\eta$ are treated as constants. 
This system  describes the evolution of a shear wave within the linearized Israel-Stewart framework (the second equation is just \eqref{Israel1} in the special case in which there is no dependence on $y$). 
If we look for solutions of the form $\{ u_y ,\Pi_{xy}\}\propto e^{i(kx-\omega t)}$, where $k\in \mathbb{R}$ and $\omega\in \mathbb{C}$ are constants, we find that $\omega$ must be a root of the polynomial $\tau_\eta\omega^2+i\omega-Dk^2$, with $D=\eta/(\rho+P)$. This produces two dispersion relations:
\begin{equation}
\label{MaestroSplinter}
\begin{split}
     \omega(k) = 
     \dfrac{i}{2\tau_\eta} \left(-1-\sqrt{1-4D\tau_\eta k^2} \, \right)  
     \,\, \Rightarrow \quad &
    \omega \approx -\dfrac{i}{\tau_\eta} + i D k^2 
     \,\,\,\, (|k|\ll k_c) \, ,
     \quad 
     \omega \approx -\dfrac{i}{2\tau_\eta} - w \, k 
     \,\,\,\,(|k|\gg k_c)
     \\
     \omega(k) = \dfrac{i}{2\tau_\eta} \left(-1+\sqrt{1-4D\tau_\eta k^2} \, \right) 
     \,\, \Rightarrow \quad &
     \, \omega \approx - i D k^2 
     \,\,\,\,\, (|k|\ll k_c) \, ,
     \quad 
     \omega \approx -\dfrac{i}{2\tau_\eta} + w \, k 
     \,\,\,\,\, \left(|k|\gg k_c\right)
     \\
\end{split} 
\end{equation} 
where $k_c=1/\sqrt{4D\tau_\eta}$ is the critical value for which the two modes coincide and $w=\sqrt{D/\tau_\eta}$ is a characteristic speed.

For large wavelengths ($|k|\ll k_c$), the first dispersion relation describes a non-hydrodynamic mode, see \cite{GavassinoQuasi2022}: even in the homogeneous limit ($k=0$) this type of perturbation relaxes exponentially in time with a timescale $\tau_\eta$. 
The second one describes a diffusive mode: for small $k$, we recover the usual Navier-Stokes-type diffusion relation $\omega \approx -iDk^2$, meaning that the shear wave is purely damped. 

Is there a regime where shear waves can propagate? In the limit of large $k$, we obtained two sound-type dispersion relations $\omega \approx \pm w k -i/(2\tau_\eta)$. Therefore, at high frequencies, the Israel-Stewart theory admits propagating shear waves, which travel at speed $w$ and decay over a timescale $2\tau_\eta$. The existence of transverse propagating waves is a distinctive feature of elastic 
(or \emph{viscoelastic}\footnote{
    Physically, a viscoelastic fluid, when subjected to a shear deformation, will produce a stress state which will eventually decay to zero, a feature that is captured by the modes in \eqref{MaestroSplinter}. The main difference with respect to a (isotropic) solid is that the viscoelastic fluid has a continuum of stress-free configurations while a viscoelastic solid may have only one \citep{CHRISTENSEN19821}, see \cite{Truesdell1965} for the same idea expressed in mathematically more rigorous terms.}) 
media. Indeed, the Maxwell model (the non-relativistic limit of the Israel-Stewart theory) is often quoted as the prototype of a viscoelastic non-Newtonian fluid \cite{CHRISTENSEN19821,Tropea_book,Li2011}. This is also consistent with the expected behaviour of liquids subject to high frequency perturbations \citep{Frenkel_book}.

\section{GENERIC-EIT dictionary}
\label{AAA}

\citet{Stricker2019} have set up a notation that is quite different from the ``standard'' EIT notation that we are adopting here. In this appendix, we show how to connect the two languages, and we prove equations \eqref{ConstitutiveGeneric} and \eqref{SecondLaw}.

First, we note that in \cite{Stricker2019} the quantity $\rho$ is the \textit{rest-mass} density, and not the energy density\footnote{
    We remove the subscripts ``f'' used by \citet{Stricker2019} to indicate that a thermodynamic quantity is measured in the fluid's local rest-frame (where $u^0=1$ and $u^j=0$), since this is a standard construction shared by most hydrodynamic frameworks. 
    }. 
Thus, what  \citet{Stricker2019} call ``$\rho$'' for us is $mn$, where $m$ is the rest mass of the constituent particles. Furthermore, they split the energy density into rest-mass plus an internal contribution. Thus, what they call ``$\rho c^2 + \epsilon$'' is exactly our $\rho$. 
To avoid confusion and maintain consistency with the main text, in the following we will keep using the symbol $\rho$ for the total energy density (in the Eckart frame) and $mn$ for the rest-mass density.

\citet{Stricker2019} introduce two structural variables, $\omega_a$ and $\alpha_{ab}$($=\alpha_{ba}$), each subject to an algebraic constraint: $\omega_a u^a =T$, $\alpha_{ab}u^a=u_b$. 
We can use the flow velocity to decompose these structural variables as follows:
\begin{equation}\label{auno}
\begin{split}
& \omega_a = \hat{\omega}_a -T u_a \spc \quad \text{with } \, \hat{\omega}_a u^a=0 \, ,\\
& \alpha_{ab}=\hat{\alpha}_{ab}-u_a u_b \spc  \text{with } \, \hat{\alpha}_{ab} u^a=\hat{\alpha}_{ba} u^a=0 \, . \\
\end{split}
\end{equation}
With the aid of the identities
\begin{equation}
\begin{split}
& \alpha^{ac}\alpha\indices{_c ^b}-\alpha^{ab}=\hat{\alpha}^{ac}\hat{\alpha}\indices{_c ^b}-\hat{\alpha}^{ab} \, , \\ & \alpha^{bc}\alpha_{bc}-1=\hat{\alpha}^{bc}\hat{\alpha}_{bc} \, , \\
\end{split}
\end{equation}
which follow directly from \eqref{auno}, we can rewrite the constitutive relations for stress-energy tensor and entropy current (as they are given in \cite{Stricker2019}) in the following form:
\begin{equation}
\begin{split}
 T^{ab} ={}& \rho u^a u^b +P h^{ab}-mn T^2 H_\omega (\hat{\omega}^a u^b+ u^a \hat{\omega}^b)-2mnT H_\alpha \hat{\alpha}^{ab}+mnT (H_\omega \hat{\omega}^a \hat{\omega}^b +2H_\alpha \hat{\alpha}^{ac}\hat{\alpha}\indices{_c ^b}) \, , \\
 s^a ={}& s u^a -mnT H_\omega \hat{\omega}^a - \dfrac{1}{2} mn (H_\omega \hat{\omega}^b \hat{\omega}_b + H_\alpha \hat{\alpha}^{bc}\hat{\alpha}_{bc}) u^a \, . \\
\end{split}
\end{equation}
To recover equation \eqref{ConstitutiveGeneric}, we only need to make the identifications below:
\begin{equation}\label{leprime4}
\begin{split}
& q^a = -mnT^2 H_\omega \hat{\omega}^a \, ,\\
& \pi^{ab} = -2mnTH_\alpha \hat{\alpha}^{ab} \, ,  \\
& b_1 = (mnT^3 H_\omega)^{-1} \, , \\
& b_2 = (4mnTH_\alpha)^{-1} \, . \\
\end{split} 
\end{equation}
Now let us focus on the entropy production rate, $\sigma:= \nabla_a s^a$. \citet{Stricker2019} provide the following formula [right-hand side of equation (13)]:
\begin{equation}\label{entropons}
\sigma = mnH_\alpha \dfrac{\overline{\alpha}^{ab}\overline{\alpha}_{ab}}{\lambda_0} + mnH_\omega \dfrac{\hat{\omega}^a \hat{\omega}_a}{\lambda_1} + mnH_\alpha \dfrac{\mathring{\alpha}^{ab}\mathring{\alpha}_{ab}}{\lambda_2} \, ,
\end{equation}
where $\overline{\alpha}_{ab}$ and $\mathring{\alpha}_{ab}$ are the irreducible parts of $\hat{\alpha}_{ab}$:
\begin{equation}
\hat{\alpha}_{ab}= \overline{\alpha}_{ab}+\mathring{\alpha}_{ab} \, ,  \spc  \overline{\alpha}_{ab} = \dfrac{1}{3}\hat{\alpha}\indices{^c _c}h_{ab} \, .
\end{equation}
With the aid of \eqref{leprime4}, we see that \eqref{entropons} is equivalent to \eqref{SecondLaw}, provided that we make the identifications
\begin{equation}\label{Leultime3}
\begin{split}
& \zeta = 4mnTH_\alpha \lambda_0 /3 \, , \\
& \kappa = mnT^2 H_\omega \lambda_1  \, ,\\
& \eta = 2mnT H_\alpha \lambda_2 \, . \\
\end{split}
\end{equation}
Equations \eqref{leprime4} and \eqref{Leultime3} fix the notation correspondence completely. 

%

\section{Equivalence of the field equations}\label{equiWuz}

In this Appendix, we prove that the field equations of the structural variables $\omega_a$ and $\alpha_{ab}$ (in the linear regime) are equivalent to the Israel-Stewart field equations \eqref{settantquattro}, \eqref{settantcinque}, and \eqref{settantasei}. 

\subsection{Some preliminary formulas}

Given our goal, it will be more convenient to work with the inverse of \eqref{leprime4} and \eqref{Leultime3}, namely
\begin{equation}\label{abra}
\hat{\omega}^a = -T b_1 q^a \, , \spc \hat{\alpha}^{ab}=-2b_2 \pi^{ab} \, ,
\end{equation}
\begin{equation}\label{cadabra}
\lambda_0 = 3 b_2 \zeta \, ,  \spc \lambda_1 = Tb_1 \kappa \, , \spc \lambda_2 = 2 b_2 \eta \, .
\end{equation}
Furthermore, keeping in mind that at equilibrium $u^a = \delta\indices{^a _0}$, we can take the variation of \eqref{auno}, and we obtain
\begin{equation}\label{alakazam}
\begin{split}
& \delta \omega_a =\delta \hat{\omega}_a -T \delta u_a-u_a \delta T  \, , \\  
& \delta\alpha_{ab}=  \delta\hat{\alpha}_{ab} -u_a \delta u_b - u_b \delta u_a \, . \\
\end{split}
\end{equation}

\subsection{Proof of the equivalence}

Let's start with equation (30) of \citet{Stricker2019}:
\begin{equation}
\partial_t \delta \alpha\indices{^j _j} - 2\partial_j \delta u^j = -\dfrac{\delta \alpha\indices{^j _j}}{\lambda_0} \, . 
\end{equation}
Using the correspondence formula $\delta \alpha\indices{^j _j}=\delta \hat{\alpha}\indices{^j _j}= -6b_2  \delta \Pi$, which follows from \eqref{abra} and \eqref{alakazam}, we immediately recover equation \eqref{settantquattro}. Now, let us focus on equation (29) of \cite{Stricker2019}:
\begin{equation}
\partial_t \delta \mathring{\alpha}_{jk} -2 \braket{\partial_j \delta u_k} = -\dfrac{\delta\mathring{\alpha}_{jk}}{\lambda_2} \, .
\end{equation}
Using the correspondence formula $\delta\mathring{\alpha}_{jk}= -2b_2 \delta \Pi_{jk}$, which follows from \eqref{abra}, we recover equation  \eqref{settantasei}. Finally, let us linearise equation (5) of \cite{Stricker2019}:
\begin{equation}
\partial_t \delta \omega_k -\partial_k \delta \omega_0 = -\dfrac{\delta \hat{\omega}_k}{\lambda_1} \, .
\end{equation}
With the aid of \eqref{alakazam}, we can rewrite it as follows:
\begin{equation}
\partial_t \bigg( \dfrac{\delta \hat{\omega}_k}{T} - \delta u_k \bigg) -\dfrac{\partial_ k \delta T}{T} = -\dfrac{\delta \hat{\omega}_k}{T\lambda_1} \, .
\end{equation}
Using the correspondence formula $\delta \hat{\omega}_k/T = -b_1 \delta q_k$, which follows from \eqref{abra}, we recover equation \eqref{settantcinque}.

\section{Thermodynamic Killing vector of a relativistic star}
\label{ThermoKilling}

The spacetime of a rotating star \cite{Gourgoulhon_intro_stars},
\begin{equation}
ds^2 = -N(r,\theta)^2 dt^2+A(r,\theta)^2(dr^2 +r^2 d\theta^2) +B(r,\theta)^2 \big[d\phi -\omega(r,\theta) dt \big]^2 \, ,
\end{equation}
admits two symmetry generators: $\partial_t$ and $\partial_\phi$. The corresponding conserved charges are the energy $U$ and the angular momentum $L$. Thus, the entropy of the heat bath can be expressed as $S_H(N_H,U_H,L_H)$, whose differential is \cite{GibbonsHawking1977}
\begin{equation}
dS_H = -\dfrac{\mu^\star}{T^\star} dN_H +\dfrac{1}{T^\star} dU_H -\dfrac{\Omega^\star}{T^\star} dL_H \, .
\end{equation}
Comparing with equation \eqref{alphastar}, we can conclude that $\alpha^\star_I=\{\mu^\star/T^\star,-1/T^\star,\Omega^\star/T^\star\}$. The conserved currents associated to $N$, $U$, and $L$ are respectively $n^a$, $- (\partial_t)_b T^{ab} $ and $(\partial_\phi)_b T^{ab}$. Hence, equation \eqref{constructphiA} becomes
\begin{equation}
\phi^a = s^a +\dfrac{\mu^\star}{T^\star} n^a +\dfrac{1}{T^\star} (\partial_t)_b T^{ab} +\dfrac{\Omega^\star}{T^\star} (\partial_\phi)_b T^{ab} \, .
\end{equation} 
Comparing this formula with equation \eqref{gringone}, we obtain
\begin{equation}
\beta^\star_b = \dfrac{1}{T^\star} (\partial_t)_b +\dfrac{\Omega^\star}{T^\star} (\partial_\phi)_b \, .
\end{equation}
As we can see, in thermodynamic equilibrium, the star is rigidly rotating, with angular velocity $\Omega^\star$ \cite{Gourgoulhon_intro_stars,Geo2020}. The quantity $T^\star$ can be interpreted as the redshifted temperature of the poles of the star, as measured at infinity. The reader should keep in mind that $T^\star$, which is a global constant, is not the local temperature $T$ [see equation \eqref{equilibrioconduzio}] that an observer would measure by bringing the fluid in direct contact with a thermometer \cite{GavassinoTermometri}. That temperature can be computed through the formula $T=(-\beta^\star_b \beta^{\star b})^{-1/2}$, and it is given by 
\begin{equation}
    T=\dfrac{T^\star}{\sqrt{N^2-B^2(\Omega^\star-\omega)^2}} \, .
\end{equation}
Finally, note that $\beta^\star_b$ is not one of the coefficients $\alpha^\star_I$ (which are always scalar constants), but it is constructed from them. The reason is that $T^{ab}$ itself is not a conserved current $J^{Ia}$, because it is not a vector field. To obtain a conserved current, we needed to contract $T^{ab}$ with the Killing vector fields of the spacetime \cite{Hawking1973}, see equation \eqref{conCurrentuzze}.

\bibliography{Biblio}

\begin{thebibliography}{95}%
\makeatletter
\providecommand \@ifxundefined [1]{%
 \@ifx{#1\undefined}
}%
\providecommand \@ifnum [1]{%
 \ifnum #1\expandafter \@firstoftwo
 \else \expandafter \@secondoftwo
 \fi
}%
\providecommand \@ifx [1]{%
 \ifx #1\expandafter \@firstoftwo
 \else \expandafter \@secondoftwo
 \fi
}%
\providecommand \natexlab [1]{#1}%
\providecommand \enquote  [1]{``#1''}%
\providecommand \bibnamefont  [1]{#1}%
\providecommand \bibfnamefont [1]{#1}%
\providecommand \citenamefont [1]{#1}%
\providecommand \href@noop [0]{\@secondoftwo}%
\providecommand \href [0]{\begingroup \@sanitize@url \@href}%
\providecommand \@href[1]{\@@startlink{#1}\@@href}%
\providecommand \@@href[1]{\endgroup#1\@@endlink}%
\providecommand \@sanitize@url [0]{\catcode `\\12\catcode `\$12\catcode
  `\&12\catcode `\#12\catcode `\^12\catcode `\_12\catcode `\%12\relax}%
\providecommand \@@startlink[1]{}%
\providecommand \@@endlink[0]{}%
\providecommand \url  [0]{\begingroup\@sanitize@url \@url }%
\providecommand \@url [1]{\endgroup\@href {#1}{\urlprefix }}%
\providecommand \urlprefix  [0]{URL }%
\providecommand \Eprint [0]{\href }%
\providecommand \doibase [0]{http://dx.doi.org/}%
\providecommand \selectlanguage [0]{\@gobble}%
\providecommand \bibinfo  [0]{\@secondoftwo}%
\providecommand \bibfield  [0]{\@secondoftwo}%
\providecommand \translation [1]{[#1]}%
\providecommand \BibitemOpen [0]{}%
\providecommand \bibitemStop [0]{}%
\providecommand \bibitemNoStop [0]{.\EOS\space}%
\providecommand \EOS [0]{\spacefactor3000\relax}%
\providecommand \BibitemShut  [1]{\csname bibitem#1\endcsname}%
\let\auto@bib@innerbib\@empty
\bibitem [{\citenamefont {{Liu}}\ \emph {et~al.}(1986)\citenamefont {{Liu}},
  \citenamefont {{M{\"u}ller}},\ and\ \citenamefont {{Ruggeri}}}]{Liu1986}%
  \BibitemOpen
  \bibfield  {author} {\bibinfo {author} {\bibfnamefont {I.~S.}\ \bibnamefont
  {{Liu}}}, \bibinfo {author} {\bibfnamefont {I.}~\bibnamefont {{M{\"u}ller}}},
  \ and\ \bibinfo {author} {\bibfnamefont {T.}~\bibnamefont {{Ruggeri}}},\
  }\href {\doibase 10.1016/0003-4916(86)90164-8} {\bibfield  {journal}
  {\bibinfo  {journal} {Annals of Physics}\ }\textbf {\bibinfo {volume}
  {169}},\ \bibinfo {pages} {191} (\bibinfo {year} {1986})}\BibitemShut
  {NoStop}%
\bibitem [{\citenamefont {Hiscock}\ and\ \citenamefont
  {Salmonson}(1991)}]{Salmonson1991}%
  \BibitemOpen
  \bibfield  {author} {\bibinfo {author} {\bibfnamefont {W.~A.}\ \bibnamefont
  {Hiscock}}\ and\ \bibinfo {author} {\bibfnamefont {J.}~\bibnamefont
  {Salmonson}},\ }\href {\doibase 10.1103/PhysRevD.43.3249} {\bibfield
  {journal} {\bibinfo  {journal} {Phys. Rev. D}\ }\textbf {\bibinfo {volume}
  {43}},\ \bibinfo {pages} {3249} (\bibinfo {year} {1991})}\BibitemShut
  {NoStop}%
\bibitem [{\citenamefont {{Rezzolla}}\ and\ \citenamefont
  {{Zanotti}}(2013)}]{rezzolla_book}%
  \BibitemOpen
  \bibfield  {author} {\bibinfo {author} {\bibfnamefont {L.}~\bibnamefont
  {{Rezzolla}}}\ and\ \bibinfo {author} {\bibfnamefont {O.}~\bibnamefont
  {{Zanotti}}},\ }\href@noop {} {\emph {\bibinfo {title} {{Relativistic
  Hydrodynamics}}}}\ (\bibinfo  {publisher} {Oxford University Press},\
  \bibinfo {year} {2013})\BibitemShut {NoStop}%
\bibitem [{\citenamefont {Stricker}\ and\ \citenamefont
  {\"Ottinger}(2019)}]{Stricker2019}%
  \BibitemOpen
  \bibfield  {author} {\bibinfo {author} {\bibfnamefont {L.}~\bibnamefont
  {Stricker}}\ and\ \bibinfo {author} {\bibfnamefont {H.~C.}\ \bibnamefont
  {\"Ottinger}},\ }\href {\doibase 10.1103/PhysRevE.99.013105} {\bibfield
  {journal} {\bibinfo  {journal} {Phys. Rev. E}\ }\textbf {\bibinfo {volume}
  {99}},\ \bibinfo {pages} {013105} (\bibinfo {year} {2019})}\BibitemShut
  {NoStop}%
\bibitem [{\citenamefont {Denicol}\ \emph {et~al.}(2012)\citenamefont
  {Denicol}, \citenamefont {Niemi}, \citenamefont {Moln\'ar},\ and\
  \citenamefont {Rischke}}]{Denicol2012Boltzmann}%
  \BibitemOpen
  \bibfield  {author} {\bibinfo {author} {\bibfnamefont {G.~S.}\ \bibnamefont
  {Denicol}}, \bibinfo {author} {\bibfnamefont {H.}~\bibnamefont {Niemi}},
  \bibinfo {author} {\bibfnamefont {E.}~\bibnamefont {Moln\'ar}}, \ and\
  \bibinfo {author} {\bibfnamefont {D.~H.}\ \bibnamefont {Rischke}},\ }\href
  {\doibase 10.1103/PhysRevD.85.114047} {\bibfield  {journal} {\bibinfo
  {journal} {Phys. Rev. D}\ }\textbf {\bibinfo {volume} {85}},\ \bibinfo
  {pages} {114047} (\bibinfo {year} {2012})}\BibitemShut {NoStop}%
\bibitem [{\citenamefont {Stewart}(1977)}]{Stewart_1977}%
  \BibitemOpen
  \bibfield  {author} {\bibinfo {author} {\bibfnamefont {J.}~\bibnamefont
  {Stewart}},\ }\href {\doibase https://doi.org/10.1098/rspa.1977.0155}
  {\bibfield  {journal} {\bibinfo  {journal} {Proceedings of the Royal Society
  A}\ }\textbf {\bibinfo {volume} {357}},\ \bibinfo {pages} {59 } (\bibinfo
  {year} {1977})}\BibitemShut {NoStop}%
\bibitem [{\citenamefont {Israel}\ and\ \citenamefont
  {Stewart}(1979)}]{Israel_Stewart_1979}%
  \BibitemOpen
  \bibfield  {author} {\bibinfo {author} {\bibfnamefont {W.}~\bibnamefont
  {Israel}}\ and\ \bibinfo {author} {\bibfnamefont {J.}~\bibnamefont
  {Stewart}},\ }\href {\doibase https://doi.org/10.1016/0003-4916(79)90130-1}
  {\bibfield  {journal} {\bibinfo  {journal} {Annals of Physics}\ }\textbf
  {\bibinfo {volume} {118}},\ \bibinfo {pages} {341 } (\bibinfo {year}
  {1979})}\BibitemShut {NoStop}%
\bibitem [{\citenamefont {{Baier}}\ \emph {et~al.}(2008)\citenamefont
  {{Baier}}, \citenamefont {{Romatschke}}, \citenamefont {{Thanh Son}},
  \citenamefont {{Starinets}},\ and\ \citenamefont {{Stephanov}}}]{Baier2008}%
  \BibitemOpen
  \bibfield  {author} {\bibinfo {author} {\bibfnamefont {R.}~\bibnamefont
  {{Baier}}}, \bibinfo {author} {\bibfnamefont {P.}~\bibnamefont
  {{Romatschke}}}, \bibinfo {author} {\bibfnamefont {D.}~\bibnamefont {{Thanh
  Son}}}, \bibinfo {author} {\bibfnamefont {A.~O.}\ \bibnamefont
  {{Starinets}}}, \ and\ \bibinfo {author} {\bibfnamefont {M.~A.}\ \bibnamefont
  {{Stephanov}}},\ }\href {\doibase 10.1088/1126-6708/2008/04/100} {\bibfield
  {journal} {\bibinfo  {journal} {Journal of High Energy Physics}\ }\textbf
  {\bibinfo {volume} {2008}},\ \bibinfo {eid} {100} (\bibinfo {year} {2008})},\
  \Eprint {http://arxiv.org/abs/0712.2451} {arXiv:0712.2451 [hep-th]}
  \BibitemShut {NoStop}%
\bibitem [{\citenamefont {{Kovtun}}\ \emph {et~al.}(2011)\citenamefont
  {{Kovtun}}, \citenamefont {{Moore}},\ and\ \citenamefont
  {{Romatschke}}}]{KovtunStickiness2011}%
  \BibitemOpen
  \bibfield  {author} {\bibinfo {author} {\bibfnamefont {P.}~\bibnamefont
  {{Kovtun}}}, \bibinfo {author} {\bibfnamefont {G.~D.}\ \bibnamefont
  {{Moore}}}, \ and\ \bibinfo {author} {\bibfnamefont {P.}~\bibnamefont
  {{Romatschke}}},\ }\href {\doibase 10.1103/PhysRevD.84.025006} {\bibfield
  {journal} {\bibinfo  {journal} {\prd}\ }\textbf {\bibinfo {volume} {84}},\
  \bibinfo {eid} {025006} (\bibinfo {year} {2011})},\ \Eprint
  {http://arxiv.org/abs/1104.1586} {arXiv:1104.1586 [hep-ph]} \BibitemShut
  {NoStop}%
\bibitem [{\citenamefont {{Grozdanov}}\ \emph {et~al.}(2019)\citenamefont
  {{Grozdanov}}, \citenamefont {{Lucas}},\ and\ \citenamefont
  {{Poovuttikul}}}]{Grozdanov2019}%
  \BibitemOpen
  \bibfield  {author} {\bibinfo {author} {\bibfnamefont {S.}~\bibnamefont
  {{Grozdanov}}}, \bibinfo {author} {\bibfnamefont {A.}~\bibnamefont
  {{Lucas}}}, \ and\ \bibinfo {author} {\bibfnamefont {N.}~\bibnamefont
  {{Poovuttikul}}},\ }\href {\doibase 10.1103/PhysRevD.99.086012} {\bibfield
  {journal} {\bibinfo  {journal} {\prd}\ }\textbf {\bibinfo {volume} {99}},\
  \bibinfo {eid} {086012} (\bibinfo {year} {2019})},\ \Eprint
  {http://arxiv.org/abs/1810.10016} {arXiv:1810.10016 [hep-th]} \BibitemShut
  {NoStop}%
\bibitem [{\citenamefont {Gavassino}\ \emph
  {et~al.}(2022{\natexlab{a}})\citenamefont {Gavassino}, \citenamefont
  {Antonelli},\ and\ \citenamefont {Haskell}}]{GavassinoQuasi2022}%
  \BibitemOpen
  \bibfield  {author} {\bibinfo {author} {\bibfnamefont {L.}~\bibnamefont
  {Gavassino}}, \bibinfo {author} {\bibfnamefont {M.}~\bibnamefont
  {Antonelli}}, \ and\ \bibinfo {author} {\bibfnamefont {B.}~\bibnamefont
  {Haskell}},\ }\href {\doibase 10.1103/PhysRevD.106.056010} {\bibfield
  {journal} {\bibinfo  {journal} {Phys. Rev. D}\ }\textbf {\bibinfo {volume}
  {106}},\ \bibinfo {pages} {056010} (\bibinfo {year}
  {2022}{\natexlab{a}})}\BibitemShut {NoStop}%
\bibitem [{\citenamefont {Jou}\ \emph {et~al.}(1999)\citenamefont {Jou},
  \citenamefont {Casas-V\'{a}zquez},\ and\ \citenamefont
  {Lebon}}]{Jou_Extended}%
  \BibitemOpen
  \bibfield  {author} {\bibinfo {author} {\bibfnamefont {D.}~\bibnamefont
  {Jou}}, \bibinfo {author} {\bibfnamefont {J.}~\bibnamefont
  {Casas-V\'{a}zquez}}, \ and\ \bibinfo {author} {\bibfnamefont
  {G.}~\bibnamefont {Lebon}},\ }\href {\doibase 10.1088/0034-4885/51/8/002}
  {\bibfield  {journal} {\bibinfo  {journal} {Reports on Progress in Physics}\
  }\textbf {\bibinfo {volume} {51}},\ \bibinfo {pages} {1105} (\bibinfo {year}
  {1999})}\BibitemShut {NoStop}%
\bibitem [{\citenamefont {Muller}\ and\ \citenamefont
  {Ruggeri}(1998)}]{Muller_book}%
  \BibitemOpen
  \bibfield  {author} {\bibinfo {author} {\bibfnamefont {I.}~\bibnamefont
  {Muller}}\ and\ \bibinfo {author} {\bibfnamefont {T.}~\bibnamefont
  {Ruggeri}},\ }\href@noop {} {\emph {\bibinfo {title} {Rational Extended
  Thermodynamics}}},\ \bibinfo {edition} {2nd}\ ed.\ (\bibinfo  {publisher}
  {Springer-Verlag New York},\ \bibinfo {year} {1998})\BibitemShut {NoStop}%
\bibitem [{\citenamefont {{Salazar}}\ and\ \citenamefont
  {{Zannias}}(2020)}]{Salazar2020}%
  \BibitemOpen
  \bibfield  {author} {\bibinfo {author} {\bibfnamefont {J.~F.}\ \bibnamefont
  {{Salazar}}}\ and\ \bibinfo {author} {\bibfnamefont {T.}~\bibnamefont
  {{Zannias}}},\ }\href {\doibase 10.1142/S0218271820300104} {\bibfield
  {journal} {\bibinfo  {journal} {International Journal of Modern Physics D}\
  }\textbf {\bibinfo {volume} {29}},\ \bibinfo {eid} {2030010} (\bibinfo {year}
  {2020})},\ \Eprint {http://arxiv.org/abs/1904.04368} {arXiv:1904.04368
  [gr-qc]} \BibitemShut {NoStop}%
\bibitem [{\citenamefont {Gavassino}\ and\ \citenamefont
  {Antonelli}(2021)}]{GavassinoFronntiers2021}%
  \BibitemOpen
  \bibfield  {author} {\bibinfo {author} {\bibfnamefont {L.}~\bibnamefont
  {Gavassino}}\ and\ \bibinfo {author} {\bibfnamefont {M.}~\bibnamefont
  {Antonelli}},\ }\href {\doibase 10.3389/fspas.2021.686344} {\bibfield
  {journal} {\bibinfo  {journal} {Front. Astron. Space Sci.}\ }\textbf
  {\bibinfo {volume} {8}},\ \bibinfo {pages} {686344} (\bibinfo {year}
  {2021})},\ \Eprint {http://arxiv.org/abs/2105.15184} {arXiv:2105.15184
  [gr-qc]} \BibitemShut {NoStop}%
\bibitem [{\citenamefont {Heller}\ \emph {et~al.}(2014)\citenamefont {Heller},
  \citenamefont {Janik}, \citenamefont {Spali\ifmmode~\acute{n}\else
  \'{n}\fi{}ski},\ and\ \citenamefont {Witaszczyk}}]{Heller2014}%
  \BibitemOpen
  \bibfield  {author} {\bibinfo {author} {\bibfnamefont {M.~P.}\ \bibnamefont
  {Heller}}, \bibinfo {author} {\bibfnamefont {R.~A.}\ \bibnamefont {Janik}},
  \bibinfo {author} {\bibfnamefont {M.}~\bibnamefont
  {Spali\ifmmode~\acute{n}\else \'{n}\fi{}ski}}, \ and\ \bibinfo {author}
  {\bibfnamefont {P.}~\bibnamefont {Witaszczyk}},\ }\href {\doibase
  10.1103/PhysRevLett.113.261601} {\bibfield  {journal} {\bibinfo  {journal}
  {Phys. Rev. Lett.}\ }\textbf {\bibinfo {volume} {113}},\ \bibinfo {pages}
  {261601} (\bibinfo {year} {2014})}\BibitemShut {NoStop}%
\bibitem [{\citenamefont {Gavassino}\ \emph {et~al.}(2021)\citenamefont
  {Gavassino}, \citenamefont {Antonelli},\ and\ \citenamefont
  {Haskell}}]{BulkGavassino}%
  \BibitemOpen
  \bibfield  {author} {\bibinfo {author} {\bibfnamefont {L.}~\bibnamefont
  {Gavassino}}, \bibinfo {author} {\bibfnamefont {M.}~\bibnamefont
  {Antonelli}}, \ and\ \bibinfo {author} {\bibfnamefont {B.}~\bibnamefont
  {Haskell}},\ }\href {\doibase 10.1088/1361-6382/abe588} {\bibfield  {journal}
  {\bibinfo  {journal} {Classical and Quantum Gravity}\ }\textbf {\bibinfo
  {volume} {38}},\ \bibinfo {pages} {075001} (\bibinfo {year}
  {2021})}\BibitemShut {NoStop}%
\bibitem [{\citenamefont {Gavassino}\ \emph
  {et~al.}(2022{\natexlab{b}})\citenamefont {Gavassino}, \citenamefont
  {Antonelli},\ and\ \citenamefont {Haskell}}]{GavassinoKhalatnikov2022}%
  \BibitemOpen
  \bibfield  {author} {\bibinfo {author} {\bibfnamefont {L.}~\bibnamefont
  {Gavassino}}, \bibinfo {author} {\bibfnamefont {M.}~\bibnamefont
  {Antonelli}}, \ and\ \bibinfo {author} {\bibfnamefont {B.}~\bibnamefont
  {Haskell}},\ }\href {\doibase 10.1103/PhysRevD.105.045011} {\bibfield
  {journal} {\bibinfo  {journal} {Phys. Rev. D}\ }\textbf {\bibinfo {volume}
  {105}},\ \bibinfo {pages} {045011} (\bibinfo {year}
  {2022}{\natexlab{b}})}\BibitemShut {NoStop}%
\bibitem [{\citenamefont {Gavassino}\ \emph
  {et~al.}(2020{\natexlab{a}})\citenamefont {Gavassino}, \citenamefont
  {Antonelli},\ and\ \citenamefont {Haskell}}]{GavassinoRadiazione}%
  \BibitemOpen
  \bibfield  {author} {\bibinfo {author} {\bibfnamefont {L.}~\bibnamefont
  {Gavassino}}, \bibinfo {author} {\bibfnamefont {M.}~\bibnamefont
  {Antonelli}}, \ and\ \bibinfo {author} {\bibfnamefont {B.}~\bibnamefont
  {Haskell}},\ }\href {\doibase 10.3390/sym12091543} {\bibfield  {journal}
  {\bibinfo  {journal} {Symmetry}\ }\textbf {\bibinfo {volume} {12}},\ \bibinfo
  {pages} {1543} (\bibinfo {year} {2020}{\natexlab{a}})}\BibitemShut {NoStop}%
\bibitem [{\citenamefont {{Camelio}}\ \emph
  {et~al.}(2022{\natexlab{a}})\citenamefont {{Camelio}}, \citenamefont
  {{Gavassino}}, \citenamefont {{Antonelli}}, \citenamefont {{Bernuzzi}},\ and\
  \citenamefont {{Haskell}}}]{Camelio2022}%
  \BibitemOpen
  \bibfield  {author} {\bibinfo {author} {\bibfnamefont {G.}~\bibnamefont
  {{Camelio}}}, \bibinfo {author} {\bibfnamefont {L.}~\bibnamefont
  {{Gavassino}}}, \bibinfo {author} {\bibfnamefont {M.}~\bibnamefont
  {{Antonelli}}}, \bibinfo {author} {\bibfnamefont {S.}~\bibnamefont
  {{Bernuzzi}}}, \ and\ \bibinfo {author} {\bibfnamefont {B.}~\bibnamefont
  {{Haskell}}},\ }\href@noop {} {\bibfield  {journal} {\bibinfo  {journal}
  {arXiv e-prints}\ ,\ \bibinfo {eid} {arXiv:2204.11809}} (\bibinfo {year}
  {2022}{\natexlab{a}})},\ \Eprint {http://arxiv.org/abs/2204.11809}
  {arXiv:2204.11809 [gr-qc]} \BibitemShut {NoStop}%
\bibitem [{\citenamefont {{Camelio}}\ \emph
  {et~al.}(2022{\natexlab{b}})\citenamefont {{Camelio}}, \citenamefont
  {{Gavassino}}, \citenamefont {{Antonelli}}, \citenamefont {{Bernuzzi}},\ and\
  \citenamefont {{Haskell}}}]{CamelioSecondo2022}%
  \BibitemOpen
  \bibfield  {author} {\bibinfo {author} {\bibfnamefont {G.}~\bibnamefont
  {{Camelio}}}, \bibinfo {author} {\bibfnamefont {L.}~\bibnamefont
  {{Gavassino}}}, \bibinfo {author} {\bibfnamefont {M.}~\bibnamefont
  {{Antonelli}}}, \bibinfo {author} {\bibfnamefont {S.}~\bibnamefont
  {{Bernuzzi}}}, \ and\ \bibinfo {author} {\bibfnamefont {B.}~\bibnamefont
  {{Haskell}}},\ }\href@noop {} {\bibfield  {journal} {\bibinfo  {journal}
  {arXiv e-prints}\ ,\ \bibinfo {eid} {arXiv:2204.11810}} (\bibinfo {year}
  {2022}{\natexlab{b}})},\ \Eprint {http://arxiv.org/abs/2204.11810}
  {arXiv:2204.11810 [gr-qc]} \BibitemShut {NoStop}%
\bibitem [{\citenamefont {Denicol}\ \emph {et~al.}(2011)\citenamefont
  {Denicol}, \citenamefont {Noronha}, \citenamefont {Niemi},\ and\
  \citenamefont {Rischke}}]{Denicol_Relaxation_2011}%
  \BibitemOpen
  \bibfield  {author} {\bibinfo {author} {\bibfnamefont {G.~S.}\ \bibnamefont
  {Denicol}}, \bibinfo {author} {\bibfnamefont {J.}~\bibnamefont {Noronha}},
  \bibinfo {author} {\bibfnamefont {H.}~\bibnamefont {Niemi}}, \ and\ \bibinfo
  {author} {\bibfnamefont {D.~H.}\ \bibnamefont {Rischke}},\ }\href {\doibase
  10.1103/PhysRevD.83.074019} {\bibfield  {journal} {\bibinfo  {journal} {Phys.
  Rev. D}\ }\textbf {\bibinfo {volume} {83}},\ \bibinfo {pages} {074019}
  (\bibinfo {year} {2011})}\BibitemShut {NoStop}%
\bibitem [{\citenamefont {Maxwell}(1867)}]{maxwell_1867}%
  \BibitemOpen
  \bibfield  {author} {\bibinfo {author} {\bibfnamefont {J.~C.}\ \bibnamefont
  {Maxwell}},\ }\href {\doibase 10.1098/rstl.1867.0004} {\bibfield  {journal}
  {\bibinfo  {journal} {Philosophical Transactions of the Royal Society of
  London}\ }\textbf {\bibinfo {volume} {157}},\ \bibinfo {pages} {49} (\bibinfo
  {year} {1867})},\ \Eprint
  {http://arxiv.org/abs/https://royalsocietypublishing.org/doi/pdf/10.1098/rstl.1867.0004}
  {https://royalsocietypublishing.org/doi/pdf/10.1098/rstl.1867.0004}
  \BibitemShut {NoStop}%
\bibitem [{\citenamefont {{Nettleton}}(1959)}]{Nettleton1959}%
  \BibitemOpen
  \bibfield  {author} {\bibinfo {author} {\bibfnamefont {R.~E.}\ \bibnamefont
  {{Nettleton}}},\ }\href {\doibase 10.1063/1.1705920} {\bibfield  {journal}
  {\bibinfo  {journal} {Physics of Fluids}\ }\textbf {\bibinfo {volume} {2}},\
  \bibinfo {pages} {256} (\bibinfo {year} {1959})}\BibitemShut {NoStop}%
\bibitem [{\citenamefont {Roylance}(2001)}]{Roylance2001}%
  \BibitemOpen
  \bibfield  {author} {\bibinfo {author} {\bibfnamefont {D.}~\bibnamefont
  {Roylance}},\ }\href
  {https://web.mit.edu/course/3/3.11/www/modules/visco.pdf} {\emph {\bibinfo
  {title} {Engineering Viscoelasticity}}}\ (\bibinfo  {publisher} {MIT lecture
  notes, Cambridge MA},\ \bibinfo {year} {2001})\BibitemShut {NoStop}%
\bibitem [{\citenamefont {Frenkel}(1955)}]{Frenkel_book}%
  \BibitemOpen
  \bibfield  {author} {\bibinfo {author} {\bibfnamefont {J.}~\bibnamefont
  {Frenkel}},\ }\href@noop {} {\emph {\bibinfo {title} {Kinetic theory of
  liquids}}},\ \bibinfo {edition} {2nd}\ ed.\ (\bibinfo  {publisher} {Dover
  Publications, New York},\ \bibinfo {year} {1955})\BibitemShut {NoStop}%
\bibitem [{\citenamefont {Andrade}\ \emph {et~al.}(2019)\citenamefont
  {Andrade}, \citenamefont {Baggioli},\ and\ \citenamefont
  {Pujol\`as}}]{BaggioliHolography2019}%
  \BibitemOpen
  \bibfield  {author} {\bibinfo {author} {\bibfnamefont {T.}~\bibnamefont
  {Andrade}}, \bibinfo {author} {\bibfnamefont {M.}~\bibnamefont {Baggioli}}, \
  and\ \bibinfo {author} {\bibfnamefont {O.}~\bibnamefont {Pujol\`as}},\ }\href
  {\doibase 10.1103/PhysRevD.100.106014} {\bibfield  {journal} {\bibinfo
  {journal} {Phys. Rev. D}\ }\textbf {\bibinfo {volume} {100}},\ \bibinfo
  {pages} {106014} (\bibinfo {year} {2019})}\BibitemShut {NoStop}%
\bibitem [{\citenamefont {Baggioli}\ \emph {et~al.}(2020)\citenamefont
  {Baggioli}, \citenamefont {Vasin}, \citenamefont {Brazhkin},\ and\
  \citenamefont {Trachenko}}]{BAGGIOLI20201}%
  \BibitemOpen
  \bibfield  {author} {\bibinfo {author} {\bibfnamefont {M.}~\bibnamefont
  {Baggioli}}, \bibinfo {author} {\bibfnamefont {M.}~\bibnamefont {Vasin}},
  \bibinfo {author} {\bibfnamefont {V.}~\bibnamefont {Brazhkin}}, \ and\
  \bibinfo {author} {\bibfnamefont {K.}~\bibnamefont {Trachenko}},\ }\href
  {\doibase https://doi.org/10.1016/j.physrep.2020.04.002} {\bibfield
  {journal} {\bibinfo  {journal} {Physics Reports}\ }\textbf {\bibinfo {volume}
  {865}},\ \bibinfo {pages} {1} (\bibinfo {year} {2020})},\ \bibinfo {note}
  {gapped momentum states}\BibitemShut {NoStop}%
\bibitem [{\citenamefont {Landau}\ and\ \citenamefont
  {Lifshitz}(1970)}]{landau7}%
  \BibitemOpen
  \bibfield  {author} {\bibinfo {author} {\bibfnamefont {L.}~\bibnamefont
  {Landau}}\ and\ \bibinfo {author} {\bibfnamefont {E.}~\bibnamefont
  {Lifshitz}},\ }\href@noop {} {\emph {\bibinfo {title} {Theory of
  elasticity}}},\ \bibinfo {number} {v. 7}\ (\bibinfo  {publisher} {Pergamon
  Press},\ \bibinfo {year} {1970})\BibitemShut {NoStop}%
\bibitem [{\citenamefont {Truesdell}\ and\ \citenamefont
  {Noll}(1965)}]{Truesdell1965}%
  \BibitemOpen
  \bibfield  {author} {\bibinfo {author} {\bibfnamefont {C.}~\bibnamefont
  {Truesdell}}\ and\ \bibinfo {author} {\bibfnamefont {W.}~\bibnamefont
  {Noll}},\ }\enquote {\bibinfo {title} {The non-linear field theories of
  mechanics},}\ \ (\bibinfo  {publisher} {Springer Berlin Heidelberg},\
  \bibinfo {year} {1965})\BibitemShut {NoStop}%
\bibitem [{\citenamefont {{Lindblom}}(1996)}]{LindblomRelaxation1996}%
  \BibitemOpen
  \bibfield  {author} {\bibinfo {author} {\bibfnamefont {L.}~\bibnamefont
  {{Lindblom}}},\ }\href {\doibase 10.1006/aphy.1996.0036} {\bibfield
  {journal} {\bibinfo  {journal} {Annals of Physics}\ }\textbf {\bibinfo
  {volume} {247}},\ \bibinfo {pages} {1} (\bibinfo {year} {1996})},\ \Eprint
  {http://arxiv.org/abs/gr-qc/9508058} {arXiv:gr-qc/9508058 [gr-qc]}
  \BibitemShut {NoStop}%
\bibitem [{\citenamefont {{Carter}}\ and\ \citenamefont
  {{Quintana}}(1972)}]{CarterQuintana1972}%
  \BibitemOpen
  \bibfield  {author} {\bibinfo {author} {\bibfnamefont {B.}~\bibnamefont
  {{Carter}}}\ and\ \bibinfo {author} {\bibfnamefont {H.}~\bibnamefont
  {{Quintana}}},\ }\href {\doibase 10.1098/rspa.1972.0164} {\bibfield
  {journal} {\bibinfo  {journal} {Proceedings of the Royal Society of London
  Series A}\ }\textbf {\bibinfo {volume} {331}},\ \bibinfo {pages} {57}
  (\bibinfo {year} {1972})}\BibitemShut {NoStop}%
\bibitem [{\citenamefont {{Trachenko}}\ and\ \citenamefont
  {{Brazhkin}}(2016)}]{Trachenko2016}%
  \BibitemOpen
  \bibfield  {author} {\bibinfo {author} {\bibfnamefont {K.}~\bibnamefont
  {{Trachenko}}}\ and\ \bibinfo {author} {\bibfnamefont {V.~V.}\ \bibnamefont
  {{Brazhkin}}},\ }\href {\doibase 10.1088/0034-4885/79/1/016502} {\bibfield
  {journal} {\bibinfo  {journal} {Reports on Progress in Physics}\ }\textbf
  {\bibinfo {volume} {79}},\ \bibinfo {eid} {016502} (\bibinfo {year}
  {2016})},\ \Eprint {http://arxiv.org/abs/1512.06592} {arXiv:1512.06592
  [cond-mat.soft]} \BibitemShut {NoStop}%
\bibitem [{\citenamefont {Grimsditch}\ \emph {et~al.}(1989)\citenamefont
  {Grimsditch}, \citenamefont {Bhadra},\ and\ \citenamefont
  {Torell}}]{Grimsditch1989}%
  \BibitemOpen
  \bibfield  {author} {\bibinfo {author} {\bibfnamefont {M.}~\bibnamefont
  {Grimsditch}}, \bibinfo {author} {\bibfnamefont {R.}~\bibnamefont {Bhadra}},
  \ and\ \bibinfo {author} {\bibfnamefont {L.~M.}\ \bibnamefont {Torell}},\
  }\href {\doibase 10.1103/PhysRevLett.62.2616} {\bibfield  {journal} {\bibinfo
   {journal} {Phys. Rev. Lett.}\ }\textbf {\bibinfo {volume} {62}},\ \bibinfo
  {pages} {2616} (\bibinfo {year} {1989})}\BibitemShut {NoStop}%
\bibitem [{\citenamefont {Scarponi}\ \emph {et~al.}(2004)\citenamefont
  {Scarponi}, \citenamefont {Comez}, \citenamefont {Fioretto},\ and\
  \citenamefont {Palmieri}}]{Scarponi2004}%
  \BibitemOpen
  \bibfield  {author} {\bibinfo {author} {\bibfnamefont {F.}~\bibnamefont
  {Scarponi}}, \bibinfo {author} {\bibfnamefont {L.}~\bibnamefont {Comez}},
  \bibinfo {author} {\bibfnamefont {D.}~\bibnamefont {Fioretto}}, \ and\
  \bibinfo {author} {\bibfnamefont {L.}~\bibnamefont {Palmieri}},\ }\href
  {\doibase 10.1103/PhysRevB.70.054203} {\bibfield  {journal} {\bibinfo
  {journal} {Phys. Rev. B}\ }\textbf {\bibinfo {volume} {70}},\ \bibinfo
  {pages} {054203} (\bibinfo {year} {2004})}\BibitemShut {NoStop}%
\bibitem [{\citenamefont {Hosokawa}\ \emph {et~al.}(2013)\citenamefont
  {Hosokawa}, \citenamefont {Munejiri}, \citenamefont {Inui}, \citenamefont
  {Kajihara}, \citenamefont {Pilgrim}, \citenamefont {Ohmasa}, \citenamefont
  {Tsutsui}, \citenamefont {Baron}, \citenamefont {Shimojo},\ and\
  \citenamefont {Hoshino}}]{Hosokawa_2013}%
  \BibitemOpen
  \bibfield  {author} {\bibinfo {author} {\bibfnamefont {S.}~\bibnamefont
  {Hosokawa}}, \bibinfo {author} {\bibfnamefont {S.}~\bibnamefont {Munejiri}},
  \bibinfo {author} {\bibfnamefont {M.}~\bibnamefont {Inui}}, \bibinfo {author}
  {\bibfnamefont {Y.}~\bibnamefont {Kajihara}}, \bibinfo {author}
  {\bibfnamefont {W.-C.}\ \bibnamefont {Pilgrim}}, \bibinfo {author}
  {\bibfnamefont {Y.}~\bibnamefont {Ohmasa}}, \bibinfo {author} {\bibfnamefont
  {S.}~\bibnamefont {Tsutsui}}, \bibinfo {author} {\bibfnamefont {A.~Q.~R.}\
  \bibnamefont {Baron}}, \bibinfo {author} {\bibfnamefont {F.}~\bibnamefont
  {Shimojo}}, \ and\ \bibinfo {author} {\bibfnamefont {K.}~\bibnamefont
  {Hoshino}},\ }\href {\doibase 10.1088/0953-8984/25/11/112101} {\bibfield
  {journal} {\bibinfo  {journal} {Journal of Physics: Condensed Matter}\
  }\textbf {\bibinfo {volume} {25}},\ \bibinfo {pages} {112101} (\bibinfo
  {year} {2013})}\BibitemShut {NoStop}%
\bibitem [{\citenamefont {Hosokawa}\ \emph {et~al.}(2015)\citenamefont
  {Hosokawa}, \citenamefont {Inui}, \citenamefont {Kajihara}, \citenamefont
  {Tsutsui},\ and\ \citenamefont {Baron}}]{Hosokawa_2015}%
  \BibitemOpen
  \bibfield  {author} {\bibinfo {author} {\bibfnamefont {S.}~\bibnamefont
  {Hosokawa}}, \bibinfo {author} {\bibfnamefont {M.}~\bibnamefont {Inui}},
  \bibinfo {author} {\bibfnamefont {Y.}~\bibnamefont {Kajihara}}, \bibinfo
  {author} {\bibfnamefont {S.}~\bibnamefont {Tsutsui}}, \ and\ \bibinfo
  {author} {\bibfnamefont {A.~Q.~R.}\ \bibnamefont {Baron}},\ }\href {\doibase
  10.1088/0953-8984/27/19/194104} {\bibfield  {journal} {\bibinfo  {journal}
  {Journal of Physics: Condensed Matter}\ }\textbf {\bibinfo {volume} {27}},\
  \bibinfo {pages} {194104} (\bibinfo {year} {2015})}\BibitemShut {NoStop}%
\bibitem [{\citenamefont {\"{O}ttinger}(1998{\natexlab{a}})}]{Otting0_1998}%
  \BibitemOpen
  \bibfield  {author} {\bibinfo {author} {\bibfnamefont {H.~C.}\ \bibnamefont
  {\"{O}ttinger}},\ }\href {\doibase
  https://doi.org/10.1016/S0378-4371(98)00045-4} {\bibfield  {journal}
  {\bibinfo  {journal} {Physica A: Statistical Mechanics and its Applications}\
  }\textbf {\bibinfo {volume} {254}},\ \bibinfo {pages} {433} (\bibinfo {year}
  {1998}{\natexlab{a}})}\BibitemShut {NoStop}%
\bibitem [{\citenamefont {\"{O}ttinger}(1998{\natexlab{b}})}]{Otting1_1998}%
  \BibitemOpen
  \bibfield  {author} {\bibinfo {author} {\bibfnamefont {H.~C.}\ \bibnamefont
  {\"{O}ttinger}},\ }\href {\doibase
  https://doi.org/10.1016/S0378-4371(98)00298-2} {\bibfield  {journal}
  {\bibinfo  {journal} {Physica A: Statistical Mechanics and its Applications}\
  }\textbf {\bibinfo {volume} {259}},\ \bibinfo {pages} {24} (\bibinfo {year}
  {1998}{\natexlab{b}})}\BibitemShut {NoStop}%
\bibitem [{\citenamefont {\"Ottinger}(1999)}]{Otting2_1999}%
  \BibitemOpen
  \bibfield  {author} {\bibinfo {author} {\bibfnamefont {H.~C.}\ \bibnamefont
  {\"Ottinger}},\ }\href {\doibase 10.1103/PhysRevD.60.103507} {\bibfield
  {journal} {\bibinfo  {journal} {Phys. Rev. D}\ }\textbf {\bibinfo {volume}
  {60}},\ \bibinfo {pages} {103507} (\bibinfo {year} {1999})}\BibitemShut
  {NoStop}%
\bibitem [{\citenamefont {Ilg}\ and\ \citenamefont
  {\"Ottinger}(1999)}]{Otting3_1999}%
  \BibitemOpen
  \bibfield  {author} {\bibinfo {author} {\bibfnamefont {P.}~\bibnamefont
  {Ilg}}\ and\ \bibinfo {author} {\bibfnamefont {H.~C.}\ \bibnamefont
  {\"Ottinger}},\ }\href {\doibase 10.1103/PhysRevD.61.023510} {\bibfield
  {journal} {\bibinfo  {journal} {Phys. Rev. D}\ }\textbf {\bibinfo {volume}
  {61}},\ \bibinfo {pages} {023510} (\bibinfo {year} {1999})}\BibitemShut
  {NoStop}%
\bibitem [{\citenamefont {Grmela}\ and\ \citenamefont
  {\"Ottinger}(1997)}]{Grmela1997}%
  \BibitemOpen
  \bibfield  {author} {\bibinfo {author} {\bibfnamefont {M.}~\bibnamefont
  {Grmela}}\ and\ \bibinfo {author} {\bibfnamefont {H.~C.}\ \bibnamefont
  {\"Ottinger}},\ }\href {\doibase 10.1103/PhysRevE.56.6620} {\bibfield
  {journal} {\bibinfo  {journal} {Phys. Rev. E}\ }\textbf {\bibinfo {volume}
  {56}},\ \bibinfo {pages} {6620} (\bibinfo {year} {1997})}\BibitemShut
  {NoStop}%
\bibitem [{\citenamefont {{{\"O}ttinger}}(2018)}]{OttingerReview2018}%
  \BibitemOpen
  \bibfield  {author} {\bibinfo {author} {\bibfnamefont {H.~C.}\ \bibnamefont
  {{{\"O}ttinger}}},\ }\href@noop {} {\bibfield  {journal} {\bibinfo  {journal}
  {arXiv e-prints}\ ,\ \bibinfo {eid} {arXiv:1810.08470}} (\bibinfo {year}
  {2018})},\ \Eprint {http://arxiv.org/abs/1810.08470} {arXiv:1810.08470
  [cond-mat.soft]} \BibitemShut {NoStop}%
\bibitem [{\citenamefont {Hiscock}\ and\ \citenamefont
  {Lindblom}(1983)}]{Hishcock1983}%
  \BibitemOpen
  \bibfield  {author} {\bibinfo {author} {\bibfnamefont {W.~A.}\ \bibnamefont
  {Hiscock}}\ and\ \bibinfo {author} {\bibfnamefont {L.}~\bibnamefont
  {Lindblom}},\ }\href {\doibase https://doi.org/10.1016/0003-4916(83)90288-9}
  {\bibfield  {journal} {\bibinfo  {journal} {Annals of Physics}\ }\textbf
  {\bibinfo {volume} {151}},\ \bibinfo {pages} {466 } (\bibinfo {year}
  {1983})}\BibitemShut {NoStop}%
\bibitem [{\citenamefont {Callen}(1985)}]{Callen_book}%
  \BibitemOpen
  \bibfield  {author} {\bibinfo {author} {\bibfnamefont {H.~B.}\ \bibnamefont
  {Callen}},\ }\href {https://cds.cern.ch/record/450289} {\emph {\bibinfo
  {title} {{Thermodynamics and an introduction to thermostatistics; 2nd
  ed.}}}}\ (\bibinfo  {publisher} {Wiley},\ \bibinfo {address} {New York, NY},\
  \bibinfo {year} {1985})\BibitemShut {NoStop}%
\bibitem [{\citenamefont {{Peliti}}(2011)}]{peliti_book}%
  \BibitemOpen
  \bibfield  {author} {\bibinfo {author} {\bibfnamefont {L.}~\bibnamefont
  {{Peliti}}},\ }\href@noop {} {\emph {\bibinfo {title} {{Statistical Mechanics
  in a Nutshell}}}},\ series ``In a nutshell''\ (\bibinfo  {publisher}
  {Princeton University Press},\ \bibinfo {year} {2011})\BibitemShut {NoStop}%
\bibitem [{\citenamefont {Bemfica}\ \emph {et~al.}(2019)\citenamefont
  {Bemfica}, \citenamefont {Disconzi},\ and\ \citenamefont
  {Noronha}}]{Bemfica2019_conformal1}%
  \BibitemOpen
  \bibfield  {author} {\bibinfo {author} {\bibfnamefont {F.~S.}\ \bibnamefont
  {Bemfica}}, \bibinfo {author} {\bibfnamefont {M.~M.}\ \bibnamefont
  {Disconzi}}, \ and\ \bibinfo {author} {\bibfnamefont {J.}~\bibnamefont
  {Noronha}},\ }\href {\doibase 10.1103/PhysRevD.100.104020} {\bibfield
  {journal} {\bibinfo  {journal} {Phys. Rev. D}\ }\textbf {\bibinfo {volume}
  {100}},\ \bibinfo {pages} {104020} (\bibinfo {year} {2019})}\BibitemShut
  {NoStop}%
\bibitem [{\citenamefont {{Kovtun}}(2019)}]{Kovtun2019}%
  \BibitemOpen
  \bibfield  {author} {\bibinfo {author} {\bibfnamefont {P.}~\bibnamefont
  {{Kovtun}}},\ }\href {\doibase 10.1007/JHEP10(2019)034} {\bibfield  {journal}
  {\bibinfo  {journal} {Journal of High Energy Physics}\ }\textbf {\bibinfo
  {volume} {2019}},\ \bibinfo {eid} {34} (\bibinfo {year} {2019})},\ \Eprint
  {http://arxiv.org/abs/1907.08191} {arXiv:1907.08191 [hep-th]} \BibitemShut
  {NoStop}%
\bibitem [{\citenamefont {Bemfica}\ \emph {et~al.}(2022)\citenamefont
  {Bemfica}, \citenamefont {Disconzi},\ and\ \citenamefont
  {Noronha}}]{BemficaDNDefinitivo2020}%
  \BibitemOpen
  \bibfield  {author} {\bibinfo {author} {\bibfnamefont {F.~S.}\ \bibnamefont
  {Bemfica}}, \bibinfo {author} {\bibfnamefont {M.~M.}\ \bibnamefont
  {Disconzi}}, \ and\ \bibinfo {author} {\bibfnamefont {J.}~\bibnamefont
  {Noronha}},\ }\href {\doibase 10.1103/PhysRevX.12.021044} {\bibfield
  {journal} {\bibinfo  {journal} {Phys. Rev. X}\ }\textbf {\bibinfo {volume}
  {12}},\ \bibinfo {pages} {021044} (\bibinfo {year} {2022})}\BibitemShut
  {NoStop}%
\bibitem [{\citenamefont {{Dore}}\ \emph {et~al.}(2022)\citenamefont {{Dore}},
  \citenamefont {{Gavassino}}, \citenamefont {{Montenegro}}, \citenamefont
  {{Shokri}},\ and\ \citenamefont {{Torrieri}}}]{DoreTorrieri2022}%
  \BibitemOpen
  \bibfield  {author} {\bibinfo {author} {\bibfnamefont {T.}~\bibnamefont
  {{Dore}}}, \bibinfo {author} {\bibfnamefont {L.}~\bibnamefont {{Gavassino}}},
  \bibinfo {author} {\bibfnamefont {D.}~\bibnamefont {{Montenegro}}}, \bibinfo
  {author} {\bibfnamefont {M.}~\bibnamefont {{Shokri}}}, \ and\ \bibinfo
  {author} {\bibfnamefont {G.}~\bibnamefont {{Torrieri}}},\ }\href {\doibase
  10.1016/j.aop.2022.168902} {\bibfield  {journal} {\bibinfo  {journal} {Annals
  of Physics}\ }\textbf {\bibinfo {volume} {442}},\ \bibinfo {eid} {168902}
  (\bibinfo {year} {2022})},\ \Eprint {http://arxiv.org/abs/2109.06389}
  {arXiv:2109.06389 [hep-th]} \BibitemShut {NoStop}%
\bibitem [{\citenamefont {{Kovtun}}(2012)}]{KovtunLecture2012}%
  \BibitemOpen
  \bibfield  {author} {\bibinfo {author} {\bibfnamefont {P.}~\bibnamefont
  {{Kovtun}}},\ }\href {\doibase 10.1088/1751-8113/45/47/473001} {\bibfield
  {journal} {\bibinfo  {journal} {Journal of Physics A Mathematical General}\
  }\textbf {\bibinfo {volume} {45}},\ \bibinfo {eid} {473001} (\bibinfo {year}
  {2012})},\ \Eprint {http://arxiv.org/abs/1205.5040} {arXiv:1205.5040
  [hep-th]} \BibitemShut {NoStop}%
\bibitem [{\citenamefont {{Becattini}}\ \emph {et~al.}(2019)\citenamefont
  {{Becattini}}, \citenamefont {{Buzzegoli}},\ and\ \citenamefont
  {{Grossi}}}]{BecattiniLecture2019}%
  \BibitemOpen
  \bibfield  {author} {\bibinfo {author} {\bibfnamefont {F.}~\bibnamefont
  {{Becattini}}}, \bibinfo {author} {\bibfnamefont {M.}~\bibnamefont
  {{Buzzegoli}}}, \ and\ \bibinfo {author} {\bibfnamefont {E.}~\bibnamefont
  {{Grossi}}},\ }\href@noop {} {\bibfield  {journal} {\bibinfo  {journal}
  {arXiv e-prints}\ ,\ \bibinfo {eid} {arXiv:1902.01089}} (\bibinfo {year}
  {2019})},\ \Eprint {http://arxiv.org/abs/1902.01089} {arXiv:1902.01089
  [cond-mat.stat-mech]} \BibitemShut {NoStop}%
\bibitem [{\citenamefont {{Romatschke}}\ and\ \citenamefont
  {{Romatschke}}(2017)}]{Romatschke2017}%
  \BibitemOpen
  \bibfield  {author} {\bibinfo {author} {\bibfnamefont {P.}~\bibnamefont
  {{Romatschke}}}\ and\ \bibinfo {author} {\bibfnamefont {U.}~\bibnamefont
  {{Romatschke}}},\ }\href@noop {} {\bibfield  {journal} {\bibinfo  {journal}
  {arXiv e-prints}\ ,\ \bibinfo {eid} {arXiv:1712.05815}} (\bibinfo {year}
  {2017})},\ \Eprint {http://arxiv.org/abs/1712.05815} {arXiv:1712.05815
  [nucl-th]} \BibitemShut {NoStop}%
\bibitem [{\citenamefont {Cattaneo}(1958)}]{cattaneo1958}%
  \BibitemOpen
  \bibfield  {author} {\bibinfo {author} {\bibfnamefont {C.}~\bibnamefont
  {Cattaneo}},\ }\href {https://books.google.pl/books?id=mHGeQwAACAAJ} {\emph
  {\bibinfo {title} {Sur une forme de l'{\'e}quation de la chaleur
  {\'e}liminant le paradoxe d'une propagation instantan{\'e}e}}},\ Comptes
  rendus hebdomadaires des s{\'e}ances de l'Acad{\'e}mie des sciences\
  (\bibinfo  {publisher} {Gauthier-Villars},\ \bibinfo {year}
  {1958})\BibitemShut {NoStop}%
\bibitem [{\citenamefont {Hiscock}\ and\ \citenamefont
  {Lindblom}(1985)}]{Hiscock_Insatibility_first_order}%
  \BibitemOpen
  \bibfield  {author} {\bibinfo {author} {\bibfnamefont {W.}~\bibnamefont
  {Hiscock}}\ and\ \bibinfo {author} {\bibfnamefont {L.}~\bibnamefont
  {Lindblom}},\ }\href {\doibase 10.1103/PhysRevD.31.725} {\bibfield  {journal}
  {\bibinfo  {journal} {Physical review D: Particles and fields}\ }\textbf
  {\bibinfo {volume} {31}},\ \bibinfo {pages} {725} (\bibinfo {year}
  {1985})}\BibitemShut {NoStop}%
\bibitem [{\citenamefont {Gavassino}(2022)}]{GavassinoSuperlum2021}%
  \BibitemOpen
  \bibfield  {author} {\bibinfo {author} {\bibfnamefont {L.}~\bibnamefont
  {Gavassino}},\ }\href {\doibase 10.1103/PhysRevX.12.041001} {\bibfield
  {journal} {\bibinfo  {journal} {Phys. Rev. X}\ }\textbf {\bibinfo {volume}
  {12}},\ \bibinfo {pages} {041001} (\bibinfo {year} {2022})}\BibitemShut
  {NoStop}%
\bibitem [{\citenamefont {Gavassino}\ \emph
  {et~al.}(2020{\natexlab{b}})\citenamefont {Gavassino}, \citenamefont
  {Antonelli},\ and\ \citenamefont {Haskell}}]{GavassinoLyapunov_2020}%
  \BibitemOpen
  \bibfield  {author} {\bibinfo {author} {\bibfnamefont {L.}~\bibnamefont
  {Gavassino}}, \bibinfo {author} {\bibfnamefont {M.}~\bibnamefont
  {Antonelli}}, \ and\ \bibinfo {author} {\bibfnamefont {B.}~\bibnamefont
  {Haskell}},\ }\href {\doibase 10.1103/physrevd.102.043018} {\bibfield
  {journal} {\bibinfo  {journal} {Physical Review D}\ }\textbf {\bibinfo
  {volume} {102}} (\bibinfo {year} {2020}{\natexlab{b}}),\
  10.1103/physrevd.102.043018}\BibitemShut {NoStop}%
\bibitem [{\citenamefont {{Gavassino}}(2021)}]{GavassinoGibbs2021}%
  \BibitemOpen
  \bibfield  {author} {\bibinfo {author} {\bibfnamefont {L.}~\bibnamefont
  {{Gavassino}}},\ }\href {\doibase 10.1088/1361-6382/ac2b0e} {\bibfield
  {journal} {\bibinfo  {journal} {Classical and Quantum Gravity}\ }\textbf
  {\bibinfo {volume} {38}},\ \bibinfo {eid} {21LT02} (\bibinfo {year}
  {2021})},\ \Eprint {http://arxiv.org/abs/2104.09142} {arXiv:2104.09142
  [gr-qc]} \BibitemShut {NoStop}%
\bibitem [{\citenamefont {{Gavassino}}\ \emph {et~al.}(2022)\citenamefont
  {{Gavassino}}, \citenamefont {{Antonelli}},\ and\ \citenamefont
  {{Haskell}}}]{GavassinoCausality2021}%
  \BibitemOpen
  \bibfield  {author} {\bibinfo {author} {\bibfnamefont {L.}~\bibnamefont
  {{Gavassino}}}, \bibinfo {author} {\bibfnamefont {M.}~\bibnamefont
  {{Antonelli}}}, \ and\ \bibinfo {author} {\bibfnamefont {B.}~\bibnamefont
  {{Haskell}}},\ }\href {\doibase 10.1103/PhysRevLett.128.010606} {\bibfield
  {journal} {\bibinfo  {journal} {\prl}\ }\textbf {\bibinfo {volume} {128}},\
  \bibinfo {eid} {010606} (\bibinfo {year} {2022})},\ \Eprint
  {http://arxiv.org/abs/2105.14621} {arXiv:2105.14621 [gr-qc]} \BibitemShut
  {NoStop}%
\bibitem [{\citenamefont
  {{Gavassino}}(2022{\natexlab{a}})}]{GavassinoStabilityCarter2022}%
  \BibitemOpen
  \bibfield  {author} {\bibinfo {author} {\bibfnamefont {L.}~\bibnamefont
  {{Gavassino}}},\ }\href {\doibase 10.1088/1361-6382/ac79f4} {\bibfield
  {journal} {\bibinfo  {journal} {Classical and Quantum Gravity}\ }\textbf
  {\bibinfo {volume} {39}},\ \bibinfo {eid} {185008} (\bibinfo {year}
  {2022}{\natexlab{a}})},\ \Eprint {http://arxiv.org/abs/2202.06760}
  {arXiv:2202.06760 [gr-qc]} \BibitemShut {NoStop}%
\bibitem [{\citenamefont {Gavassino}(2020)}]{GavassinoTermometri}%
  \BibitemOpen
  \bibfield  {author} {\bibinfo {author} {\bibfnamefont {L.}~\bibnamefont
  {Gavassino}},\ }\href {\doibase 10.1007/s10701-020-00393-x} {\bibfield
  {journal} {\bibinfo  {journal} {Found. Phys.}\ }\textbf {\bibinfo {volume}
  {50}},\ \bibinfo {pages} {1554} (\bibinfo {year} {2020})},\ \Eprint
  {http://arxiv.org/abs/2005.06396} {arXiv:2005.06396 [gr-qc]} \BibitemShut
  {NoStop}%
\bibitem [{\citenamefont {{Misner}}\ \emph {et~al.}(1973)\citenamefont
  {{Misner}}, \citenamefont {{Thorne}},\ and\ \citenamefont
  {{Wheeler}}}]{MTW_book}%
  \BibitemOpen
  \bibfield  {author} {\bibinfo {author} {\bibfnamefont {C.~W.}\ \bibnamefont
  {{Misner}}}, \bibinfo {author} {\bibfnamefont {K.~S.}\ \bibnamefont
  {{Thorne}}}, \ and\ \bibinfo {author} {\bibfnamefont {J.~A.}\ \bibnamefont
  {{Wheeler}}},\ }\href@noop {} {\emph {\bibinfo {title} {San Francisco:
  W.H.~Freeman and Co., 1973}}}\ (\bibinfo {year} {1973})\BibitemShut {NoStop}%
\bibitem [{\citenamefont {{Weinberg}}(1972)}]{Weinberg_book_1972}%
  \BibitemOpen
  \bibfield  {author} {\bibinfo {author} {\bibfnamefont {S.}~\bibnamefont
  {{Weinberg}}},\ }\href@noop {} {\emph {\bibinfo {title} {{Gravitation and
  Cosmology: Principles and Applications of the General Theory of
  Relativity}}}}\ (\bibinfo {year} {1972})\BibitemShut {NoStop}%
\bibitem [{\citenamefont
  {{Gavassino}}(2022{\natexlab{b}})}]{GavassinoLorentzInvariance2022}%
  \BibitemOpen
  \bibfield  {author} {\bibinfo {author} {\bibfnamefont {L.}~\bibnamefont
  {{Gavassino}}},\ }\href {\doibase 10.1007/s10701-021-00518-w} {\bibfield
  {journal} {\bibinfo  {journal} {Foundations of Physics}\ }\textbf {\bibinfo
  {volume} {52}},\ \bibinfo {eid} {11} (\bibinfo {year}
  {2022}{\natexlab{b}})},\ \Eprint {http://arxiv.org/abs/2105.09294}
  {arXiv:2105.09294 [gr-qc]} \BibitemShut {NoStop}%
\bibitem [{\citenamefont {Huang}(1987)}]{huang_book}%
  \BibitemOpen
  \bibfield  {author} {\bibinfo {author} {\bibfnamefont {K.}~\bibnamefont
  {Huang}},\ }\href@noop {} {\emph {\bibinfo {title} {Statistical
  Mechanics}}},\ \bibinfo {edition} {2nd}\ ed.\ (\bibinfo  {publisher} {John
  Wiley \& Sons},\ \bibinfo {year} {1987})\BibitemShut {NoStop}%
\bibitem [{\citenamefont {{Gavassino}}\ and\ \citenamefont
  {{Antonelli}}(2020)}]{Termo}%
  \BibitemOpen
  \bibfield  {author} {\bibinfo {author} {\bibfnamefont {L.}~\bibnamefont
  {{Gavassino}}}\ and\ \bibinfo {author} {\bibfnamefont {M.}~\bibnamefont
  {{Antonelli}}},\ }\href {\doibase 10.1088/1361-6382/ab5f23} {\bibfield
  {journal} {\bibinfo  {journal} {Classical and Quantum Gravity}\ }\textbf
  {\bibinfo {volume} {37}},\ \bibinfo {eid} {025014} (\bibinfo {year}
  {2020})},\ \Eprint {http://arxiv.org/abs/1906.03140} {arXiv:1906.03140
  [gr-qc]} \BibitemShut {NoStop}%
\bibitem [{\citenamefont {Landau}\ and\ \citenamefont
  {Lifshitz}(2013)}]{landau5}%
  \BibitemOpen
  \bibfield  {author} {\bibinfo {author} {\bibfnamefont {L.}~\bibnamefont
  {Landau}}\ and\ \bibinfo {author} {\bibfnamefont {E.}~\bibnamefont
  {Lifshitz}},\ }\href {https://books.google.pl/books?id=VzgJN-XPTRsC} {\emph
  {\bibinfo {title} {Statistical Physics}}},\ \bibinfo {number} {v. 5}\
  (\bibinfo  {publisher} {Elsevier Science},\ \bibinfo {year}
  {2013})\BibitemShut {NoStop}%
\bibitem [{\citenamefont {Stueckelberg}(1962)}]{Stuekelberg1962}%
  \BibitemOpen
  \bibfield  {author} {\bibinfo {author} {\bibfnamefont {E.}~\bibnamefont
  {Stueckelberg}},\ }\href@noop {} {\bibfield  {journal} {\bibinfo  {journal}
  {Helvetica Physica Acta}\ }\textbf {\bibinfo {volume} {35}} (\bibinfo {year}
  {1962})}\BibitemShut {NoStop}%
\bibitem [{\citenamefont {Israel}(2009)}]{Israel_2009_inbook}%
  \BibitemOpen
  \bibfield  {author} {\bibinfo {author} {\bibfnamefont {W.}~\bibnamefont
  {Israel}},\ }\enquote {\bibinfo {title} {Relativistic thermodynamics},}\ in\
  \href {\doibase 10.1007/978-3-7643-8878-2_8} {\emph {\bibinfo {booktitle}
  {E.C.G. Stueckelberg, An Unconventional Figure of Twentieth Century Physics:
  Selected Scientific Papers with Commentaries}}},\ \bibinfo {editor} {edited
  by\ \bibinfo {editor} {\bibfnamefont {J.}~\bibnamefont {Lacki}}, \bibinfo
  {editor} {\bibfnamefont {H.}~\bibnamefont {Ruegg}}, \ and\ \bibinfo {editor}
  {\bibfnamefont {G.}~\bibnamefont {Wanders}}}\ (\bibinfo  {publisher}
  {Birkh{\"a}user Basel},\ \bibinfo {address} {Basel},\ \bibinfo {year}
  {2009})\ pp.\ \bibinfo {pages} {101--113}\BibitemShut {NoStop}%
\bibitem [{\citenamefont {Hawking}\ and\ \citenamefont
  {Ellis}(2011)}]{Hawking1973}%
  \BibitemOpen
  \bibfield  {author} {\bibinfo {author} {\bibfnamefont {S.~W.}\ \bibnamefont
  {Hawking}}\ and\ \bibinfo {author} {\bibfnamefont {G.~F.~R.}\ \bibnamefont
  {Ellis}},\ }\href {\doibase 10.1017/CBO9780511524646} {\emph {\bibinfo
  {title} {{The Large Scale Structure of Space-Time}}}},\ Cambridge Monographs
  on Mathematical Physics\ (\bibinfo  {publisher} {Cambridge University
  Press},\ \bibinfo {year} {2011})\BibitemShut {NoStop}%
\bibitem [{\citenamefont {Pathria}\ and\ \citenamefont
  {Beale}(2011)}]{Pathria2011}%
  \BibitemOpen
  \bibfield  {author} {\bibinfo {author} {\bibfnamefont {R.}~\bibnamefont
  {Pathria}}\ and\ \bibinfo {author} {\bibfnamefont {P.~D.}\ \bibnamefont
  {Beale}},\ }in\ \href {\doibase
  https://doi.org/10.1016/B978-0-12-382188-1.00015-3} {\emph {\bibinfo
  {booktitle} {Statistical Mechanics (Third Edition)}}},\ \bibinfo {editor}
  {edited by\ \bibinfo {editor} {\bibfnamefont {R.}~\bibnamefont {Pathria}}\
  and\ \bibinfo {editor} {\bibfnamefont {P.~D.}\ \bibnamefont {Beale}}}\
  (\bibinfo  {publisher} {Academic Press},\ \bibinfo {address} {Boston},\
  \bibinfo {year} {2011})\ \bibinfo {edition} {third edition}\ ed.,\ pp.\
  \bibinfo {pages} {583--635}\BibitemShut {NoStop}%
\bibitem [{\citenamefont {{Becattini}}(2016)}]{BecattiniBeta2016}%
  \BibitemOpen
  \bibfield  {author} {\bibinfo {author} {\bibfnamefont {F.}~\bibnamefont
  {{Becattini}}},\ }\href {\doibase 10.5506/APhysPolB.47.1819} {\bibfield
  {journal} {\bibinfo  {journal} {Acta Physica Polonica B}\ }\textbf {\bibinfo
  {volume} {47}},\ \bibinfo {pages} {1819} (\bibinfo {year} {2016})},\ \Eprint
  {http://arxiv.org/abs/1606.06605} {arXiv:1606.06605 [gr-qc]} \BibitemShut
  {NoStop}%
\bibitem [{\citenamefont {Gibbons}\ and\ \citenamefont
  {Hawking}(1977)}]{GibbonsHawking1977}%
  \BibitemOpen
  \bibfield  {author} {\bibinfo {author} {\bibfnamefont {G.~W.}\ \bibnamefont
  {Gibbons}}\ and\ \bibinfo {author} {\bibfnamefont {S.~W.}\ \bibnamefont
  {Hawking}},\ }\href {\doibase 10.1103/PhysRevD.15.2752} {\bibfield  {journal}
  {\bibinfo  {journal} {Phys. Rev. D}\ }\textbf {\bibinfo {volume} {15}},\
  \bibinfo {pages} {2752} (\bibinfo {year} {1977})}\BibitemShut {NoStop}%
\bibitem [{\citenamefont {Almaalol}\ \emph {et~al.}(2022)\citenamefont
  {Almaalol}, \citenamefont {Dore},\ and\ \citenamefont
  {Noronha-Hostler}}]{Almaalol2022}%
  \BibitemOpen
  \bibfield  {author} {\bibinfo {author} {\bibfnamefont {D.}~\bibnamefont
  {Almaalol}}, \bibinfo {author} {\bibfnamefont {T.}~\bibnamefont {Dore}}, \
  and\ \bibinfo {author} {\bibfnamefont {J.}~\bibnamefont {Noronha-Hostler}},\
  }\href@noop {} {\  (\bibinfo {year} {2022})},\ \Eprint
  {http://arxiv.org/abs/2209.11210} {arXiv:2209.11210 [hep-th]} \BibitemShut
  {NoStop}%
\bibitem [{\citenamefont {{Gavassino}}\ \emph {et~al.}(2023)\citenamefont
  {{Gavassino}}, \citenamefont {{Disconzi}},\ and\ \citenamefont
  {{Noronha}}}]{GavassinoUniversality2023}%
  \BibitemOpen
  \bibfield  {author} {\bibinfo {author} {\bibfnamefont {L.}~\bibnamefont
  {{Gavassino}}}, \bibinfo {author} {\bibfnamefont {M.~M.}\ \bibnamefont
  {{Disconzi}}}, \ and\ \bibinfo {author} {\bibfnamefont {J.}~\bibnamefont
  {{Noronha}}},\ }\href {\doibase 10.48550/arXiv.2302.03478} {\bibfield
  {journal} {\bibinfo  {journal} {arXiv e-prints}\ ,\ \bibinfo {eid}
  {arXiv:2302.03478}} (\bibinfo {year} {2023})},\ \Eprint
  {http://arxiv.org/abs/2302.03478} {arXiv:2302.03478 [nucl-th]} \BibitemShut
  {NoStop}%
\bibitem [{\citenamefont {LaSalle}\ and\ \citenamefont
  {Lefschetz}(1961)}]{lasalle1961stability}%
  \BibitemOpen
  \bibfield  {author} {\bibinfo {author} {\bibfnamefont {J.}~\bibnamefont
  {LaSalle}}\ and\ \bibinfo {author} {\bibfnamefont {S.}~\bibnamefont
  {Lefschetz}},\ }\href {https://books.google.pl/books?id=fAZRAAAAMAAJ} {\emph
  {\bibinfo {title} {Stability by Liapunov's Direct Method: With
  Applications}}},\ Mathematics in science andengineering, v.4\ (\bibinfo
  {publisher} {Academic Press},\ \bibinfo {year} {1961})\BibitemShut {NoStop}%
\bibitem [{\citenamefont {{Gavassino}}(2023)}]{GavassinoBounds2023}%
  \BibitemOpen
  \bibfield  {author} {\bibinfo {author} {\bibfnamefont {L.}~\bibnamefont
  {{Gavassino}}},\ }\href {\doibase 10.48550/arXiv.2301.06651} {\bibfield
  {journal} {\bibinfo  {journal} {arXiv e-prints}\ ,\ \bibinfo {eid}
  {arXiv:2301.06651}} (\bibinfo {year} {2023})},\ \Eprint
  {http://arxiv.org/abs/2301.06651} {arXiv:2301.06651 [hep-th]} \BibitemShut
  {NoStop}%
\bibitem [{\citenamefont {Geroch}\ and\ \citenamefont
  {Lindblom}(1991)}]{Geroch_Lindblom_1991_causal}%
  \BibitemOpen
  \bibfield  {author} {\bibinfo {author} {\bibfnamefont {R.}~\bibnamefont
  {Geroch}}\ and\ \bibinfo {author} {\bibfnamefont {L.}~\bibnamefont
  {Lindblom}},\ }\href {\doibase https://doi.org/10.1016/0003-4916(91)90063-E}
  {\bibfield  {journal} {\bibinfo  {journal} {Annals of Physics}\ }\textbf
  {\bibinfo {volume} {207}},\ \bibinfo {pages} {394 } (\bibinfo {year}
  {1991})}\BibitemShut {NoStop}%
\bibitem [{\citenamefont {{Pu}}\ \emph {et~al.}(2010)\citenamefont {{Pu}},
  \citenamefont {{Koide}},\ and\ \citenamefont {{Rischke}}}]{Pu2010}%
  \BibitemOpen
  \bibfield  {author} {\bibinfo {author} {\bibfnamefont {S.}~\bibnamefont
  {{Pu}}}, \bibinfo {author} {\bibfnamefont {T.}~\bibnamefont {{Koide}}}, \
  and\ \bibinfo {author} {\bibfnamefont {D.~H.}\ \bibnamefont {{Rischke}}},\
  }\href {\doibase 10.1103/PhysRevD.81.114039} {\bibfield  {journal} {\bibinfo
  {journal} {\prd}\ }\textbf {\bibinfo {volume} {81}},\ \bibinfo {eid} {114039}
  (\bibinfo {year} {2010})},\ \Eprint {http://arxiv.org/abs/0907.3906}
  {arXiv:0907.3906 [hep-ph]} \BibitemShut {NoStop}%
\bibitem [{\citenamefont {{Bemfica}}\ \emph {et~al.}(2019)\citenamefont
  {{Bemfica}}, \citenamefont {{Disconzi}},\ and\ \citenamefont
  {{Noronha}}}]{Causality_bulk}%
  \BibitemOpen
  \bibfield  {author} {\bibinfo {author} {\bibfnamefont {F.~S.}\ \bibnamefont
  {{Bemfica}}}, \bibinfo {author} {\bibfnamefont {M.~M.}\ \bibnamefont
  {{Disconzi}}}, \ and\ \bibinfo {author} {\bibfnamefont {J.}~\bibnamefont
  {{Noronha}}},\ }\href@noop {} {\bibfield  {journal} {\bibinfo  {journal}
  {arXiv e-prints}\ ,\ \bibinfo {eid} {arXiv:1901.06701}} (\bibinfo {year}
  {2019})},\ \Eprint {http://arxiv.org/abs/1901.06701} {arXiv:1901.06701
  [gr-qc]} \BibitemShut {NoStop}%
\bibitem [{\citenamefont {{Geroch}}\ and\ \citenamefont
  {{Lindblom}}(1990)}]{GerochLindblom1990}%
  \BibitemOpen
  \bibfield  {author} {\bibinfo {author} {\bibfnamefont {R.}~\bibnamefont
  {{Geroch}}}\ and\ \bibinfo {author} {\bibfnamefont {L.}~\bibnamefont
  {{Lindblom}}},\ }\href {\doibase 10.1103/PhysRevD.41.1855} {\bibfield
  {journal} {\bibinfo  {journal} {\prd}\ }\textbf {\bibinfo {volume} {41}},\
  \bibinfo {pages} {1855} (\bibinfo {year} {1990})}\BibitemShut {NoStop}%
\bibitem [{\citenamefont {Priou}(1991)}]{PriouCOMPAR1991}%
  \BibitemOpen
  \bibfield  {author} {\bibinfo {author} {\bibfnamefont {D.}~\bibnamefont
  {Priou}},\ }\href {\doibase 10.1103/PhysRevD.43.1223} {\bibfield  {journal}
  {\bibinfo  {journal} {Phys. Rev. D}\ }\textbf {\bibinfo {volume} {43}},\
  \bibinfo {pages} {1223} (\bibinfo {year} {1991})}\BibitemShut {NoStop}%
\bibitem [{\citenamefont {{Schmitt}}\ and\ \citenamefont
  {{Shternin}}(2018)}]{Schmitt2018ASSL}%
  \BibitemOpen
  \bibfield  {author} {\bibinfo {author} {\bibfnamefont {A.}~\bibnamefont
  {{Schmitt}}}\ and\ \bibinfo {author} {\bibfnamefont {P.}~\bibnamefont
  {{Shternin}}},\ }in\ \href {\doibase 10.1007/978-3-319-97616-7_9} {\emph
  {\bibinfo {booktitle} {Astrophysics and Space Science Library}}},\ \bibinfo
  {series} {Astrophysics and Space Science Library}, Vol.\ \bibinfo {volume}
  {457},\ \bibinfo {editor} {edited by\ \bibinfo {editor} {\bibfnamefont
  {L.}~\bibnamefont {{Rezzolla}}}, \bibinfo {editor} {\bibfnamefont
  {P.}~\bibnamefont {{Pizzochero}}}, \bibinfo {editor} {\bibfnamefont {D.~I.}\
  \bibnamefont {{Jones}}}, \bibinfo {editor} {\bibfnamefont {N.}~\bibnamefont
  {{Rea}}}, \ and\ \bibinfo {editor} {\bibfnamefont {I.}~\bibnamefont
  {{Vida{\~n}a}}}}\ (\bibinfo {year} {2018})\ p.\ \bibinfo {pages} {455},\
  \Eprint {http://arxiv.org/abs/1711.06520} {arXiv:1711.06520 [astro-ph.HE]}
  \BibitemShut {NoStop}%
\bibitem [{\citenamefont {Shternin}\ and\ \citenamefont
  {Yakovlev}(2008)}]{Shternin_core_08}%
  \BibitemOpen
  \bibfield  {author} {\bibinfo {author} {\bibfnamefont {P.~S.}\ \bibnamefont
  {Shternin}}\ and\ \bibinfo {author} {\bibfnamefont {D.~G.}\ \bibnamefont
  {Yakovlev}},\ }\href {\doibase 10.1103/PhysRevD.78.063006} {\bibfield
  {journal} {\bibinfo  {journal} {Phys. Rev. D}\ }\textbf {\bibinfo {volume}
  {78}},\ \bibinfo {pages} {063006} (\bibinfo {year} {2008})}\BibitemShut
  {NoStop}%
\bibitem [{\citenamefont {Escobedo}\ \emph {et~al.}(2009)\citenamefont
  {Escobedo}, \citenamefont {Mannarelli},\ and\ \citenamefont
  {Manuel}}]{EscobedoManuel2009}%
  \BibitemOpen
  \bibfield  {author} {\bibinfo {author} {\bibfnamefont {M.~A.}\ \bibnamefont
  {Escobedo}}, \bibinfo {author} {\bibfnamefont {M.}~\bibnamefont
  {Mannarelli}}, \ and\ \bibinfo {author} {\bibfnamefont {C.}~\bibnamefont
  {Manuel}},\ }\href {\doibase 10.1103/PhysRevA.79.063623} {\bibfield
  {journal} {\bibinfo  {journal} {Phys. Rev. A}\ }\textbf {\bibinfo {volume}
  {79}},\ \bibinfo {pages} {063623} (\bibinfo {year} {2009})}\BibitemShut
  {NoStop}%
\bibitem [{\citenamefont {{Manuel}}\ and\ \citenamefont
  {{Tolos}}(2021)}]{Manuel2021Univ}%
  \BibitemOpen
  \bibfield  {author} {\bibinfo {author} {\bibfnamefont {C.}~\bibnamefont
  {{Manuel}}}\ and\ \bibinfo {author} {\bibfnamefont {L.}~\bibnamefont
  {{Tolos}}},\ }\href {\doibase 10.3390/universe7030059} {\bibfield  {journal}
  {\bibinfo  {journal} {Universe}\ }\textbf {\bibinfo {volume} {7}},\ \bibinfo
  {pages} {59} (\bibinfo {year} {2021})},\ \Eprint
  {http://arxiv.org/abs/2101.09000} {arXiv:2101.09000 [nucl-th]} \BibitemShut
  {NoStop}%
\bibitem [{\citenamefont {Kovtun}\ and\ \citenamefont
  {Starinets}(2005)}]{KovtunHolography2005}%
  \BibitemOpen
  \bibfield  {author} {\bibinfo {author} {\bibfnamefont {P.~K.}\ \bibnamefont
  {Kovtun}}\ and\ \bibinfo {author} {\bibfnamefont {A.~O.}\ \bibnamefont
  {Starinets}},\ }\href {\doibase 10.1103/PhysRevD.72.086009} {\bibfield
  {journal} {\bibinfo  {journal} {Phys. Rev. D}\ }\textbf {\bibinfo {volume}
  {72}},\ \bibinfo {pages} {086009} (\bibinfo {year} {2005})}\BibitemShut
  {NoStop}%
\bibitem [{\citenamefont {{Noronha}}\ and\ \citenamefont
  {{Denicol}}(2011)}]{Noronha2011}%
  \BibitemOpen
  \bibfield  {author} {\bibinfo {author} {\bibfnamefont {J.}~\bibnamefont
  {{Noronha}}}\ and\ \bibinfo {author} {\bibfnamefont {G.~S.}\ \bibnamefont
  {{Denicol}}},\ }\href@noop {} {\bibfield  {journal} {\bibinfo  {journal}
  {arXiv e-prints}\ ,\ \bibinfo {eid} {arXiv:1104.2415}} (\bibinfo {year}
  {2011})},\ \Eprint {http://arxiv.org/abs/1104.2415} {arXiv:1104.2415
  [hep-th]} \BibitemShut {NoStop}%
\bibitem [{\citenamefont {{Rougemont}}\ \emph {et~al.}(2021)\citenamefont
  {{Rougemont}}, \citenamefont {{Noronha}}, \citenamefont {{Barreto}},
  \citenamefont {{Denicol}},\ and\ \citenamefont {{Dore}}}]{Rougemont2021}%
  \BibitemOpen
  \bibfield  {author} {\bibinfo {author} {\bibfnamefont {R.}~\bibnamefont
  {{Rougemont}}}, \bibinfo {author} {\bibfnamefont {J.}~\bibnamefont
  {{Noronha}}}, \bibinfo {author} {\bibfnamefont {W.}~\bibnamefont
  {{Barreto}}}, \bibinfo {author} {\bibfnamefont {G.~S.}\ \bibnamefont
  {{Denicol}}}, \ and\ \bibinfo {author} {\bibfnamefont {T.}~\bibnamefont
  {{Dore}}},\ }\href {\doibase 10.1103/PhysRevD.104.126012} {\bibfield
  {journal} {\bibinfo  {journal} {\prd}\ }\textbf {\bibinfo {volume} {104}},\
  \bibinfo {eid} {126012} (\bibinfo {year} {2021})},\ \Eprint
  {http://arxiv.org/abs/2105.02378} {arXiv:2105.02378 [nucl-th]} \BibitemShut
  {NoStop}%
\bibitem [{\citenamefont
  {{Gavassino}}(2022{\natexlab{c}})}]{GavassinoCasimir2022}%
  \BibitemOpen
  \bibfield  {author} {\bibinfo {author} {\bibfnamefont {L.}~\bibnamefont
  {{Gavassino}}},\ }\href@noop {} {\bibfield  {journal} {\bibinfo  {journal}
  {arXiv e-prints}\ ,\ \bibinfo {eid} {arXiv:2210.05067}} (\bibinfo {year}
  {2022}{\natexlab{c}})},\ \Eprint {http://arxiv.org/abs/2210.05067}
  {arXiv:2210.05067 [nucl-th]} \BibitemShut {NoStop}%
\bibitem [{\citenamefont {Christensen}(1982)}]{CHRISTENSEN19821}%
  \BibitemOpen
  \bibfield  {author} {\bibinfo {author} {\bibfnamefont {R.}~\bibnamefont
  {Christensen}},\ }\href {\doibase
  https://doi.org/10.1016/B978-0-12-174252-2.50005-3} {\emph {\bibinfo {title}
  {Theory of Viscoelasticity}}},\ \bibinfo {edition} {2nd}\ ed.\ (\bibinfo
  {publisher} {Academic Press},\ \bibinfo {year} {1982})\BibitemShut {NoStop}%
\bibitem [{\citenamefont {Tropea}\ \emph {et~al.}(2007)\citenamefont {Tropea},
  \citenamefont {Yarin},\ and\ \citenamefont {Foss}}]{Tropea_book}%
  \BibitemOpen
  \bibfield  {author} {\bibinfo {author} {\bibfnamefont {C.}~\bibnamefont
  {Tropea}}, \bibinfo {author} {\bibfnamefont {A.}~\bibnamefont {Yarin}}, \
  and\ \bibinfo {author} {\bibfnamefont {J.}~\bibnamefont {Foss}},\ }\href@noop
  {} {\emph {\bibinfo {title} {Springer Handbook of Experimental Fluid
  Mechanics}}}\ (\bibinfo  {publisher} {Springer},\ \bibinfo {year}
  {2007})\BibitemShut {NoStop}%
\bibitem [{\citenamefont {Li}\ \emph {et~al.}(2011)\citenamefont {Li},
  \citenamefont {Luo}, \citenamefont {Qi},\ and\ \citenamefont
  {Zhang}}]{Li2011}%
  \BibitemOpen
  \bibfield  {author} {\bibinfo {author} {\bibfnamefont {X.~K.}\ \bibnamefont
  {Li}}, \bibinfo {author} {\bibfnamefont {Y.}~\bibnamefont {Luo}}, \bibinfo
  {author} {\bibfnamefont {Y.}~\bibnamefont {Qi}}, \ and\ \bibinfo {author}
  {\bibfnamefont {R.}~\bibnamefont {Zhang}},\ }\href {\doibase
  https://doi.org/10.1016/j.apm.2010.11.003} {\bibfield  {journal} {\bibinfo
  {journal} {Applied Mathematical Modelling}\ }\textbf {\bibinfo {volume}
  {35}},\ \bibinfo {pages} {2309} (\bibinfo {year} {2011})}\BibitemShut
  {NoStop}%
\bibitem [{\citenamefont {{Gourgoulhon}}(2010)}]{Gourgoulhon_intro_stars}%
  \BibitemOpen
  \bibfield  {author} {\bibinfo {author} {\bibfnamefont {E.}~\bibnamefont
  {{Gourgoulhon}}},\ }\href@noop {} {\bibfield  {journal} {\bibinfo  {journal}
  {arXiv e-prints}\ ,\ \bibinfo {eid} {arXiv:1003.5015}} (\bibinfo {year}
  {2010})},\ \Eprint {http://arxiv.org/abs/1003.5015} {arXiv:1003.5015 [gr-qc]}
  \BibitemShut {NoStop}%
\bibitem [{\citenamefont {{Gavassino}}\ \emph {et~al.}(2020)\citenamefont
  {{Gavassino}}, \citenamefont {{Antonelli}}, \citenamefont {{Pizzochero}},\
  and\ \citenamefont {{Haskell}}}]{Geo2020}%
  \BibitemOpen
  \bibfield  {author} {\bibinfo {author} {\bibfnamefont {L.}~\bibnamefont
  {{Gavassino}}}, \bibinfo {author} {\bibfnamefont {M.}~\bibnamefont
  {{Antonelli}}}, \bibinfo {author} {\bibfnamefont {P.~M.}\ \bibnamefont
  {{Pizzochero}}}, \ and\ \bibinfo {author} {\bibfnamefont {B.}~\bibnamefont
  {{Haskell}}},\ }\href {\doibase 10.1093/mnras/staa886} {\bibfield  {journal}
  {\bibinfo  {journal} {\mnras}\ }\textbf {\bibinfo {volume} {494}},\ \bibinfo
  {pages} {3562} (\bibinfo {year} {2020})},\ \Eprint
  {http://arxiv.org/abs/2001.08951} {arXiv:2001.08951 [astro-ph.HE]}
  \BibitemShut {NoStop}%
\end{thebibliography}%

\label{lastpage}

\end{document}